\documentclass[useA MS,usenatbib]{mn2e}

\usepackage{lscape}
\usepackage{tabularx}
\usepackage{xtab}
\usepackage{graphicx}
\usepackage{placeins}
\usepackage{amsfonts,amsmath,amssymb}
\usepackage{url}
\usepackage{color}
\usepackage{times}

\usepackage{enumerate}
\include{JournalAbbr}    
\newcommand{\msun}{~\mathrm{M}_{\odot}}

\title[Star formation in the Illustris simulation]{The star formation main sequence and stellar mass assembly of galaxies in the Illustris simulation}

\author[Sparre et al.]{\parbox[t]{\textwidth}{
		Martin Sparre$^{1}$\thanks{E-mail:sparre@dark-cosmology.dk},
		Christopher C. Hayward$^{2,3}$\thanks{Moore Prize Postdoctoral Scholar in Theoretical Astrophysics},
		Volker Springel$^{3,4}$,
                Mark Vogelsberger$^5$,
                Shy Genel$^6$,
                Paul Torrey$^{2,5,6}$,
                Dylan Nelson$^6$,
                Debora Sijacki$^7$,
                Lars Hernquist$^6$
		\vspace*{6pt}} \\
$^{1}$Dark Cosmology Centre, Niels Bohr Institute, University of Copenhagen, Juliane Maries Vej 30, 2100 Copenhagen, Denmark\\
$^2$TAPIR, Mailcode 350-17, California Institute of Technology, Pasadena, CA 91125, USA\\
$^3$Heidelberger Institut f{\"u}r Theoretische Studien, Schloss-Wolfsbrunnenweg 35, 69118 Heidelberg, Germany\\
$^4$Zentrum f\"ur Astronomie der Universit\"at Heidelberg, Astronomisches Recheninstitut, M\"onchhofstrasse 12-14, 69120 Heidelberg, Germany\\
$^5$Department of Physics, Kavli Institute for Astrophysics and Space Research, Massachusetts Institute of Technology, Cambridge, MA 02139, USA\\
$^6$Harvard-Smithsonian Center for Astrophysics, 60 Garden Street, Cambridge, MA 02138, USA\\
$^7$Kavli Institute for Cosmology, Cambridge, and Institute of Astronomy, Madingley Road, Cambridge, CB3 0HA, UK
}

\begin{document}

\date{\today}

\pagerange{\pageref{firstpage}--\pageref{lastpage}} \pubyear{2014}

\maketitle

\label{firstpage}

\begin{abstract}
  Understanding the physical processes that drive star formation is a key
  challenge for galaxy formation models. In this article we study the
  tight correlation between the star formation rate (SFR) and stellar mass
  of galaxies at a given redshift, how halo growth influences star formation,
  and star formation histories of individual galaxies. We study these topics 
  using Illustris, a state-of-the-art cosmological hydrodynamical simulation
  of galaxy formation.
  Illustris reproduces the observed relation (the star formation main sequence; SFMS)
  between SFR and stellar mass at redshifts $z=0$ and $z=4$,
  but at intermediate redshifts of $z\simeq 1-2$, the simulated SFMS
  has a significantly lower normalisation than reported by observations.
  The scatter in the relation is consistent with
  the observed scatter. However, the fraction of outliers above the
  SFR-stellar mass relation in Illustris is less than that observed.
  Galaxies with halo masses of $\sim 10^{12} \msun$ dominate the SFR
  density of the Universe, in agreement with the results of abundance
  matching. Furthermore, more-massive galaxies tend to form the bulk
  of their stars at high redshift, which indicates that `downsizing'
  occurs in Illustris.  We also studied the star formation histories
  of individual galaxies, including the use of a principal component analysis decomposition.
  We find that for fixed stellar mass, galaxies that form
  earlier have more-massive black holes at $z=0$, indicating that star
  formation and black hole growth are tightly linked processes in
  Illustris. While many of the properties of normal star-forming
  galaxies are well-reproduced in the Illustris simulation, forming a
  realistic population of starbursts will likely require higher
  resolution and probably a more sophisticated treatment of star
  formation and feedback from stars and black holes.
\end{abstract}

\section{Introduction}

In the $\Lambda$CDM paradigm, galaxies reside in dark matter halos
that are built up hierarchically as gravity amplifies perturbations
created in the early Universe \citep{1970A&A.....5...84Z}. The
structure and substructure of dark matter halos, and the cosmic web
surrounding them, have been intensively studied using numerical
simulations \citep[e.g.][]{2005Natur.435..629S, 2008Natur.454..735D,
  2011ApJ...740..102K}, in which the dark matter is modelled as
collisionless particles that interact with each other only through
gravity. In the last decade, such simulations have led to a general
consensus about the distribution of dark matter on large
scales. However, the formation and evolution of the baryons embedded
in these halos are far from understood. In the canonical theory of
galaxy formation
\citep[e.g.][]{1977ApJ...211..638S,1977MNRAS.179..541R,1978MNRAS.183..341W},
galaxies form stars when hot gas radiates away energy, cools and
looses pressure support. The evolution of such galaxies is further influenced
by their merger history, accretion of gas and dark matter, and
regulation of star formation by feedback processes related to stellar
winds, supernovae and active galactic nuclei. Much of our understanding of galaxies is based on observed relations and
physical modelling of galaxy structure. Important observations include
the relation between luminosity and velocity widths of galaxies
\citep{1976ApJ...204..668F,1977A&A....54..661T}, the global star
formation rate as a function of redshift
\citep{1996ApJ...460L...1L,1998ApJ...498..106M}, relations between
mass and metallicity \citep{2004ApJ...613..898T,2010MNRAS.408.2115M},
global star formation laws
\citep{1959ApJ...129..243S,1998ApJ...498..541K}, and the morphologies
of galaxies \citep{1926ApJ....64..321H,1980ApJ...236..351D}.

A recently established relation is the so-called `star formation main
sequence' (SFMS), which is an approximately linear
relation between the star formation rate (SFR) and the stellar mass ($M_*$) of star-forming galaxies.
The relation exists at both low \citep[$z<1$;][]{2004MNRAS.351.1151B, 2007ApJS..173..267S} and high
\citep[$z\gtrsim 1$;][]{2007ApJ...670..156D} redshift and is recovered in optical \citep{2014arXiv1411.5687T}, infrared
\citep{2011A&A...533A.119E} and radio observations \citep{2011ApJ...730...61K}. It is a 
tight relation in the sense that the scatter around the relation is small (e.g. \citealt{2014arXiv1405.2041S}
reports a scatter of $\sigma \simeq 0.2$ dex).
The normalisation of the SFMS is
observed to increase from $z=0$ to $z=2$, the redshift at which the
global star formation rate density peaks.  The tightness of the SFMS
and the fact that most star-forming galaxies lie on it imply that the
bulk of the star formation in the Universe occurs in a quasi-steady
state \citep{Noeske2007b} and that the fraction of a given
star-forming galaxy's lifetime during which it lies significantly above the SFMS
because of e.g. merger-induced starbursts is small.\footnote{It is
  sometimes claimed that galaxies on the SFMS must not be undergoing
  mergers.  This is a misconception: for most of the duration of
  mergers, the SFR is not elevated significantly by the interaction
  \citep[e.g.][]{Cox2008,2010MNRAS.402.1693H}. Consequently, merging
  galaxies often lie on the SFMS (see fig. 11 of
  \citealt{Hayward2012}; \citealt{2014MNRAS.443L..49P} also
  finds major mergers to be on the SFMS during most of the merger process}).
  Thus, it is important to not equate galaxies
  that lie above the SFMS with mergers. Instead, galaxies above the
  SFMS should be referred to as `starbursts' (by definition), which
  may or may not be merger-induced. Galaxies on the SFMS should be
  considered `quiescently star-forming', and such galaxies may still
  be involved in an ongoing merger.

In addition to characterising star-forming galaxies, the SFMS also
provides a natural way to define starbursts as galaxies with SFRs well
above the SFMS value for their stellar mass and redshift
\citep{2011ApJ...739L..40R,2012ApJ...747L..31S,2014arXiv1406.4132A}. Despite
having large SFRs compared with normal galaxies, starbursts account
for only a small fraction ($\sim 5-10$ per cent) of the global SFR
density \citep{2011ApJ...739L..40R} because they are rare and
short-lived (because of their short gas-consumption timescales;
\citealt{2009ApJ...698.1437K,2010MNRAS.407.2091G,2010ApJ...714L.118D}). This
minor contribution of starbursts to the total SFR density is
consistent with semi-empirical models for infrared galaxy number
counts \citep{2012ApJ...757L..23B} and luminosity functions
\citep{2010MNRAS.402.1693H}.  `Red and dead' or quiescent galaxies are
galaxies that lie significantly below the SFMS.  These galaxies are
typically elliptical galaxies \citep{2011ApJ...742...96W} with little
gas available for star formation. They are likely the descendants of
starbursts after their intense star formation has been quenched by
feedback from active galactic nuclei \citep[AGN;
e.g.][]{1988ApJ...325...74S,2014ApJ...782...68T,2014arXiv1406.0002C}
or other processes. Because
we are interested in actively star-forming galaxies, we will largely
ignore the quiescent galaxy population in this work.

Several attempts have been made to reproduce the SFMS in
hydrodynamical simulations of galaxies
\citep{Dave2011,2014MNRAS.437.3529K,2014MNRAS.438.1985T} and (semi-)analytical
models \citep{2010MNRAS.405.1690D, Dave2012,Dekel2013}.
With both methods, a tight relationship between SFR and stellar mass can be
recovered. However, reproducing the evolution of the normalisation is
a challenge for theoretical models \citep{2008MNRAS.385..147D,
  2009ApJ...705..617D}. The main problem is producing the correct
normalisation at both $z=0$ and $z=2$. Potential solutions to this
problem have been suggested, including a varying IMF
\citep{2008MNRAS.385..147D} and modification of the timescale for
reincorporation of gas ejected by feedback processes
\citep{2014arXiv1403.1585M}.

A different important characteristic of a population of galaxies is the connection between the growth of halos and the formation of stars inside them. Dark matter halos build up hierarchically through accretion and mergers. The formation of stars is a more complex phenomenon that is heavily influenced by feedback processes and gas cooling. By matching the abundance of halos in cosmological dark matter simulations to real observations of galaxies it has been shown that stars form most efficiently in $\simeq 10^{12}\msun$ halos \citep{2013MNRAS.428.3121M,2013ApJ...762L..31B, 2013ApJ...770...57B,2014arXiv1401.7329K}, which is believed to occur because star formation is suppressed by stellar feedback and feedback from active galactic nuclei at lower and higher halo masses, respectively \citep{2013MNRAS.436.3031V,2014MNRAS.438.1985T,2014arXiv1407.7040S}. An implication is that galaxies that reside in massive (e.g. $\sim10^{14}\msun$ at $z=0$) halos formed their stars earlier than galaxies that reside in $\sim10^{12}\msun$ halos (at $z=0$) because galaxies inside $10^{12}\msun$ halos at $z=0$ are still forming stars with the highest possible efficiency. A consequence of this complicated relation between halo growth and star formation is the `downsizing' scenario, in which the galaxies with the most-massive stellar components (e.g. $M_*\sim10^{11}\msun$) at $z=0$ formed their stars earlier than galaxies of more-moderate masses (e.g. $M_*\sim10^{10}\msun$).

The aim of this article is to study properties of star-forming
galaxies, especially the SFMS and the relation between halo growth and star formation, in the Illustris cosmological
simulation.  Section~\ref{Sec:IllustrisSim} describes Illustris,
Section~\ref{Sec:StarformingPopulation} analyses properties of the
star formation main sequence, and Section~\ref{Section:GlobalSFR} examines
how the halo mass affects star formation in galaxies. In Section~\ref{Sec:500Sample} we study star formation histories of galaxies, and we examine how the star formation history of a galaxy depends on its dark matter halo mass and black hole mass. In Section~\ref{WhatDrivesScatter}, we use the results from the previous sections to study the origin of the scatter in the star formation main sequence. We
discuss our findings in Section~\ref{Discussion} and summarize our
conclusions in Section~\ref{sec:conclusions}.

\begin{figure*}
\centering
\includegraphics[width = 0.98 \textwidth]{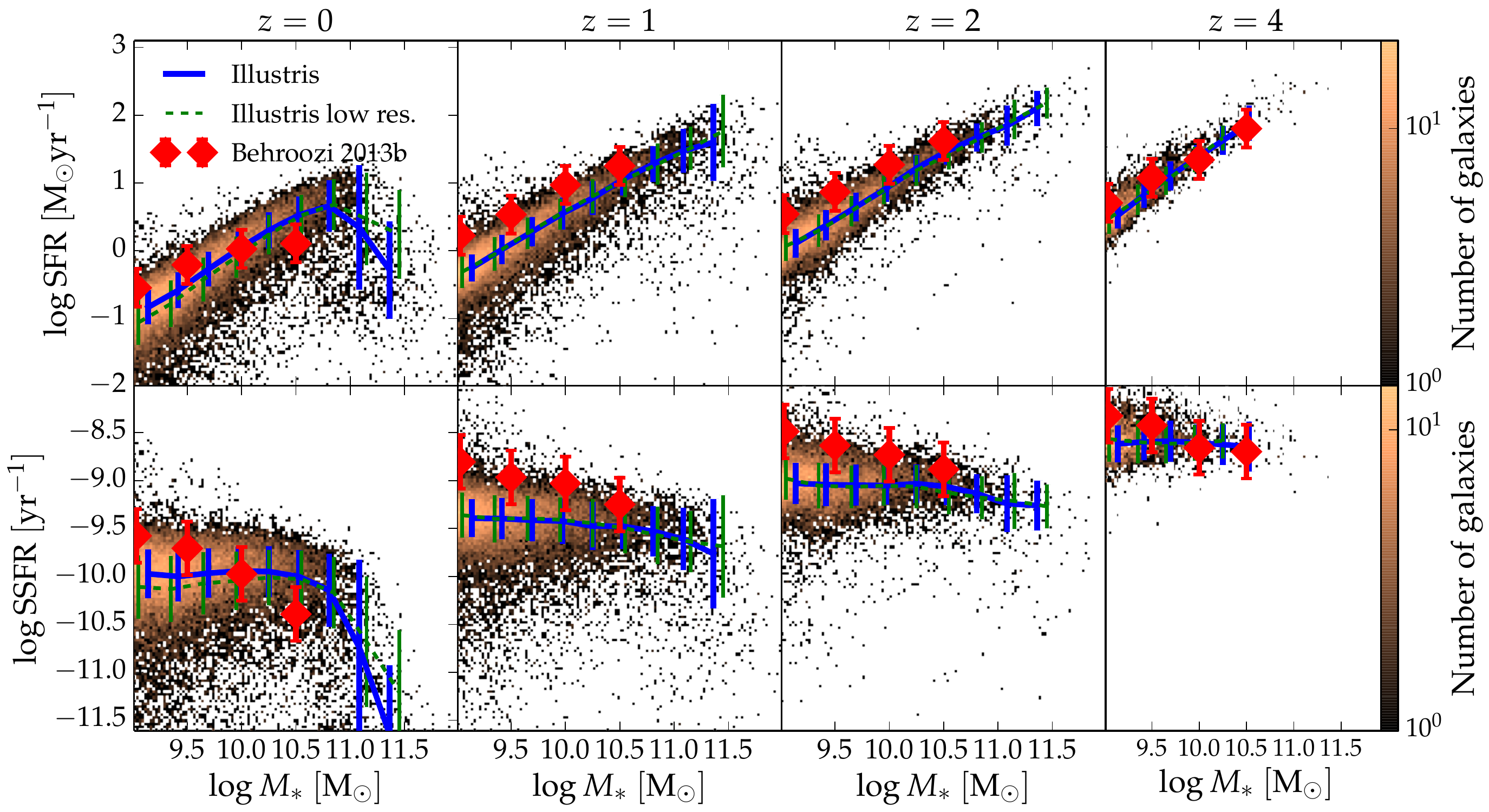}
\caption{The star formation main sequence at $z=0$, $1$, $2$ and $4$ for the
  Illustris simulation (the two-dimensional histogram in the background). The upper panels show the median SFR, and the lower
  panels show the median SSFR. The simulation results are compared to the
  compilation of observations from \citet{2013ApJ...770...57B}, where
  the error bars indicate $68\%$ confidence intervals for the inter-publication variance. The solid and dashed lines indicate the median 
  relation from the low-resolution and high-resolution versions of Illustris, respectively, and the associated error bars denote the $1\sigma$ errors estimated by fitting Gaussian functions to the SFR distributions of galaxies in narrow mass bins.}
\label{MainSequenceEvolution}
\end{figure*}

\section{The Illustris simulation}\label{Sec:IllustrisSim}

Illustris is a cosmological hydrodynamical simulation of a comoving volume
of $(106.5 \,\text{Mpc})^3$. The
cosmological model used in Illustris is the $\Lambda$CDM
cosmology with parameters from the WMAP7 data release
\citep{2013ApJS..208...19H}. Besides gravity and hydrodynamics, it includes
treatments of gas cooling, star formation, and feedback from stellar winds,
supernovae and AGN \citep[see][]{2013MNRAS.436.3031V,
2014arXiv1405.2921V,2014Natur.509..177V,2014MNRAS.438.1985T}.
The Illustris simulation has previously been used to study a range of
different galaxy properties, such as the evolution of damped
Ly-$\alpha$ absorbers \citep{2014arXiv1405.3994B}, the formation of elliptical and spiral
galaxies \citep{2014Natur.509..177V}, the evolution of galaxy properties from high to low redshift \citep{2014arXiv1405.3749G}, and the relation between the dark
matter and stellar components in the faint outskirts of galaxies
\citep{2014arXiv1406.1174P}.

The initial conditions at a redshift of $z=127$ include $1820^3$ dark
matter particles, $1820^3$ gas cells, and $1820^3$ Monte
Carlo tracer particles \citep{2013MNRAS.435.1426G}. The gravitational
softening length for the dark matter particles is 1420 comoving pc. The gravitational softening lengths of the gas cells depend on the cell size; they have a minimum value of 710 physical pc. Star particles have a softening length of 1420 comoving pc at $z\ge 1$, and at lower redshifts, the softening length is fixed to 710 physical pc. The mass of the dark
matter particles is $6.26\times 10^{6} \msun$, and the gas cell
target mass is $1.26\times 10^{6} \msun$. In this paper we will also
study a simulation with lower resolution (a total of $3\times 910^3$
dark matter, gas and tracer particles, implying 8 times worse mass
resolution and two times poorer spatial resolution), but with the same
physical galaxy formation model. We will refer to this simulation as
the \emph{Illustris low resolution run}.

The hydrodynamical calculations in Illustris are done with the {\small
  AREPO} code \citep{2010MNRAS.401..791S}, where the hydrodynamical
forces are computed on a moving mesh built with a Voronoi
tessellation. With a spatial resolution of $\simeq 1$ kpc there is no
hope to resolve giant molecular clouds, where star formation takes
place in the Universe. Instead a sub-resolution model is implemented
\citep{2003MNRAS.339..289S,2013MNRAS.436.3031V,2014arXiv1405.2921V},
where unresolved physical processes such as the formation of molecular
clouds, thermal instabilities, and supernova feedback are
coarsely described with an effective equation of state. When a gas
cell exceeds a hydrogen number density of $\rho_\text{th} = 0.13$ cm$^{-3}$ it
produces star particles stochastically on a density-dependent
timescale of
\begin{align} 
t_* (\rho )= t_0^* \left( \frac{\rho}{\rho_\text{th}}\right)^{-1/2},
\end{align}
where $t_0^*=2.2$ Gyr. With the chosen values of $\rho_\text{th}$ and
$t_0^*$, galaxies obey the empirical Kennicutt-Schmidt relation between
the gas surface density and the star formation rate per surface area
of a galaxy \citep{1989ApJ...344..685K}. A star particle in this model
represents an entire stellar population born with a Chabrier initial
mass function \citep{2003PASP..115..763C}. During each timestep of the
simulation, the amount of H, He, C, N, O, Ne, Mg, Si and Fe released by
each stellar population is calculated and returned to the gas.

The formation of stars is accompanied by the release of kinetic winds
from supernovae, which contribute to expelling the surrounding gas and
to the chemical enrichment of the interstellar gas. The wind velocity
is 3.7 times the one-dimensional velocity dispersion of the dark
matter near the star forming region.

In friends-of-friends groups more massive than $1.7\times 10^{10}\msun$ black holes are seeded, and a model of AGN feedback is included \citep{2005MNRAS.361..776S,
  2007MNRAS.380..877S}, where a quasar can be in a radio-quiet or
radio-loud mode. In the latter mode, thermal energy is injected into
the gas surrounding the black hole. Also included is a treatment of AGN radiative feedback, which heats the gas surrounding the AGN and changes its ionization state \citep[see full description in][]{2013MNRAS.436.3031V}.

Also implemented are processes
such as radiative cooling, chemical evolution, and an ultraviolet
background. For a full description of the physical model of Illustris,
see \citet{2013MNRAS.436.3031V} and \citet{2014MNRAS.438.1985T}.

\section{The star formation main sequence}\label{Sec:StarformingPopulation}

A \emph{star formation main sequence} (SFMS) that relates the star
formation rate and the stellar mass of galaxies is recovered in
Illustris; see Figure~\ref{MainSequenceEvolution}, which shows the
relation at $z=0$, $1$, $2$ and $4$. The SFMS is plotted in terms
of both the SFR and specific SFR, $\text{SSFR}\equiv
\text{SFR}/M_*$, versus $M_*$. The simulation is compared with the compilation of
observations by \citet[][Table~8]{2013ApJ...770...57B}, who fit a
relation to a large number of measurements from different authors and
quantified the scatter in the observations of the SFMS from
different publications. This scatter, which is denoted by the error bars
in Figure~\ref{MainSequenceEvolution}, quantifies the inter-publication variance of the SFMS and therefore accounts for the systematic error arising when measuring the SFMS with different
methods. The normalisation in Illustris agrees with the
observational constraints from \citet{2013ApJ...770...57B} at $z=0$.
The figure also shows that the normalisation of the SFMS in Illustris is well converged above $M_*=10^{9}\msun$.

\begin{figure}
\centering
\includegraphics[width = 0.45 \textwidth]{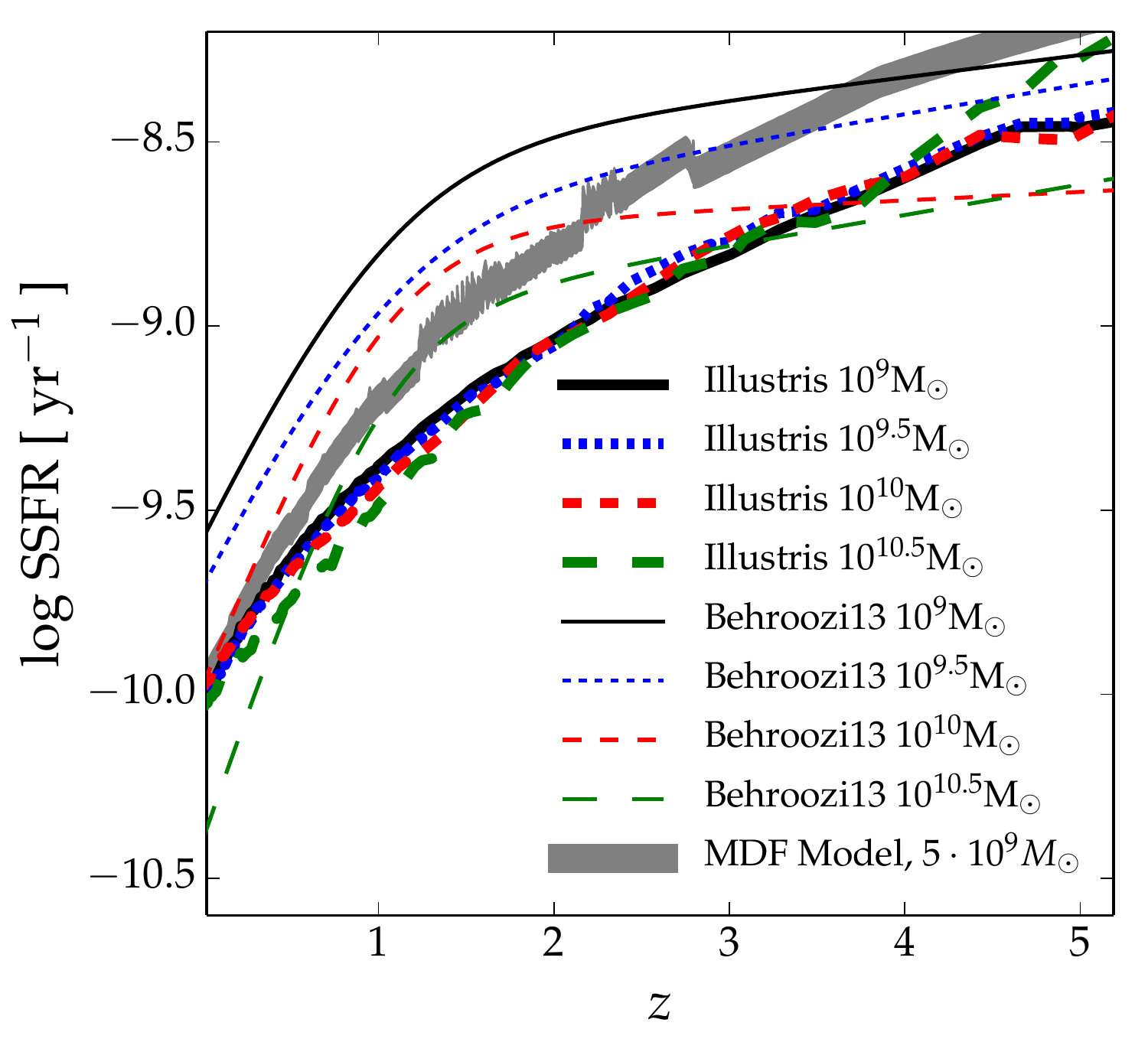}
\caption{The median SSFR as function of redshift for galaxies with
  different stellar masses. At each redshift the SSFR of galaxies in
  the given stellar mass bin is computed. The thin lines show the
  compilation of observations from \citet{2013ApJ...770...57B}, and the thick lines show the evolution in Illustris. The \emph{gray shaded} regions show the analytical model from \citet{2014arXiv1411.1157M}. The
  different evolution of the SSFR from $z=4$ to $z=0$ is closely
  related to the too low normalisation of the star formation main
  sequence at $z=1$ and $z=2$ in Figure~\ref{MainSequenceEvolution}.}
\label{Fig:SSFR_Redshift}
\end{figure}

At $z=4$, Illustris is also in good agreement with the observed
relation. However, despite being in good agreement at $z=0$ and $z=4$,
the normalisation of the SFMS is significantly lower than the
observational constraints at intermediate redshifts of $z=1$ and
$z=2$ (this has been previously noted and discussed for Illustris; see \citealt{2014arXiv1405.3749G}). Several studies have
previously pointed out discrepancies between the observed SFMS
relation and galaxy formation models
\citep[e.g.,][]{2007ApJ...670..156D, 2008MNRAS.385..147D,
  2009ApJ...705..617D}, especially at $z\simeq 2$.

The SSFR in Illustris becomes approximately independent of mass for
$M_*<10^{10.5} \msun$ (Figure~\ref{MainSequenceEvolution},
\emph{lower panels}), which is a small but remarkable difference from
the observations of \citet{2013ApJ...770...57B}, which indicate a declining SSFR as function of
mass. Figure~\ref{Fig:SSFR_Redshift} shows the redshift evolution of
the SSFR for galaxies with different stellar masses. The SSFR is here
determined by calculating the normalisation of the SFMS in different
stellar mass bins. Since the SSFR
is independent of mass at fixed redshift in Illustris, the SSFRs of
galaxies from different mass bins have the same redshift
evolution. The fitting relations from \citet{2013ApJ...770...57B} show
a somewhat different evolution, partially because the SSFR is
mass-dependent at fixed redshift. We note that the problem with
reproducing the evolution of the SSFR (in
Figure~\ref{Fig:SSFR_Redshift}) is closely related to the problem of
reproducing the normalisation of the SFMS at $z=1$ and $z=2$ (in
Figure~\ref{MainSequenceEvolution}). The different behaviour of SSFR$(z)$ in Illustris and observations also shows that the good match of the SFMS to observations at $z\simeq 4$ is most likely a coincidence.

The tensions between the observed and simulated SSFR are consistent with the lack of a sharp cut-off of the stellar mass function in
Illustris \citep{2014arXiv1405.2921V,2014arXiv1405.3749G} and the star formation efficiency peak in Illustris \citep{2014arXiv1405.2921V} being broader than that derived with abundance matching by \citet{2013ApJ...770...57B}. \citet{2014arXiv1405.3749G} also studied the redshift evolution of the SSFR in Illustris and found that the SSFR of galaxies is closely tied to the galaxies' dark matter accretion rate. 
In our model, the velocities of the stellar winds scale in proportion to the local dark matter velocity dispersion. Therefore, a natural explanation for the tensions outlined above is that the star formation and feedback processes are too closely linked to the dark matter evolution in Illustris.

In Figure~\ref{Fig:SSFR_Redshift}, we also compare our results with the analytical model from \citet{2014arXiv1411.1157M}. This model relies on an analytical approximation of halo growth from a cosmological simulation \citep{2009Natur.457..451D}. The star formation rate of a galaxy associated to a halo is determined from the equilibrium assumption that the gas mass in the ISM is non-evolving. The baryon accretion rate, mass loading of stellar winds, feedback processes that prevent gas from entering a galaxy and the timescale for reincorporating gas ejected by feedback are all closely tied to the dark matter growth rate. The eight free parameters used to control these feedback processes have been fit with a Monte Carlo Markov Chain method to maximize the agreement with the SFMS, mass-metallicity relation and halo-mass$-$stellar-mass relation at $z=0,1$ and 2. This model is one of many similar analytical galaxy formation models \citep{2010ApJ...718.1001B,2012MNRAS.421...98D,2013ApJ...772..119L,2014MNRAS.444.2071D}. As in Illustris, the SSFR at $z<2$ is under-predicted by the  \citet{2014arXiv1411.1157M} model. This supports our conclusion that it is necessary to untie the feedback processes from the dark matter growth rate in order to reproduce the observed evolution of the SSFR of galaxies. 
\citet{2014arXiv1411.1157M} matches the observations slightly better than Illustris. This is likely because it is computationally cheaper to tune model parameters in an analytical model than in a cosmological simulation.

\begin{figure}
\centering
\includegraphics[width = 0.45\textwidth]{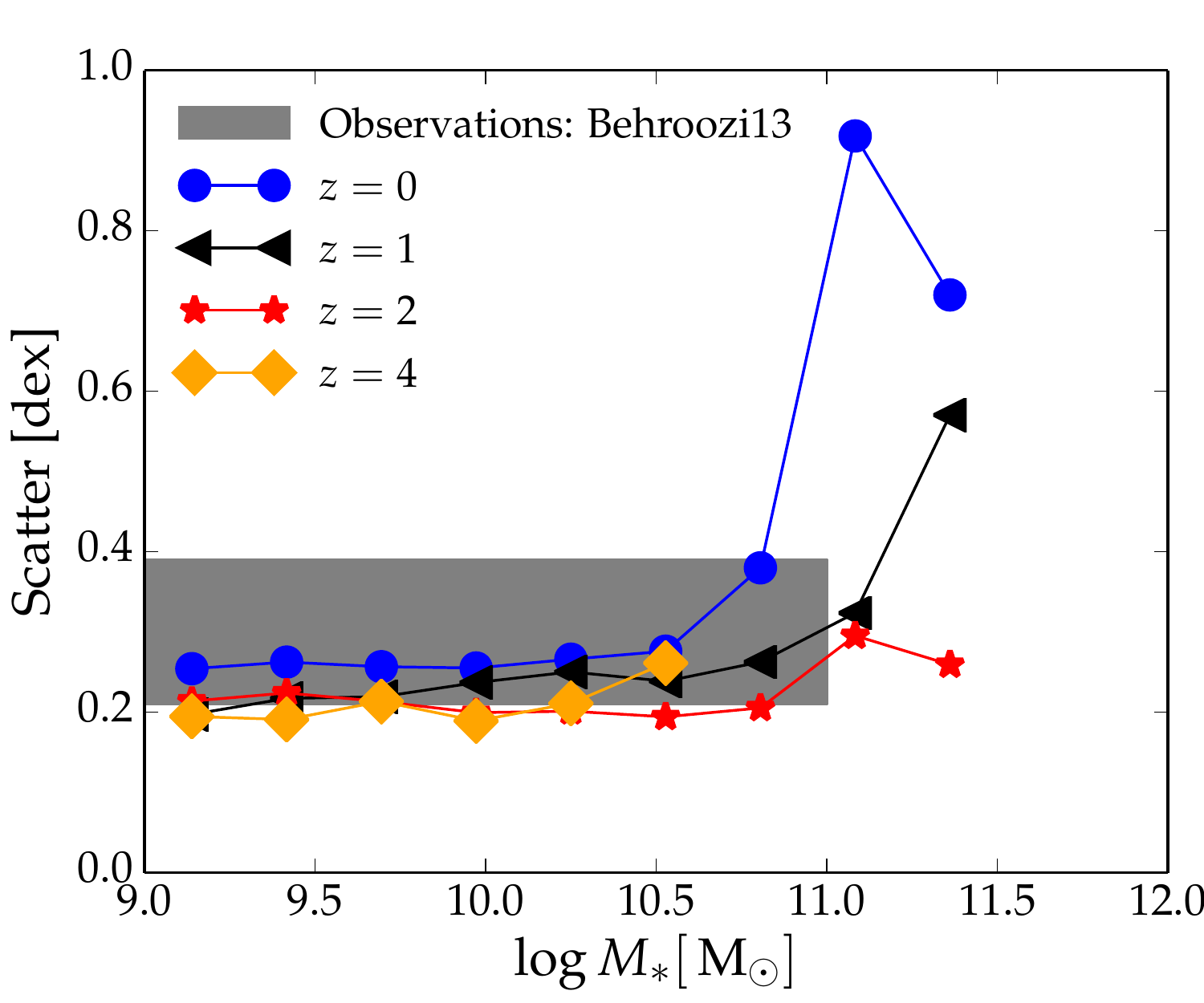}
\caption{The scatter in the main sequence as function of mass for four
  different redshifts. The grey box shows the range of scatter in
  typical observations (from \citealt{2013ApJ...770...57B}). Above $10^{11}\msun$ there are no reliable
  observational constraints for the scatter.}
\label{Fig82_MSScatter}
\end{figure}

\subsection{The scatter in the main sequence} \label{Sec:ScatterInMS}

The intrinsic scatter in the galaxy main sequence is predicted to be
driven by the different gas accretion histories of different galaxies
\citep[see the analytical modelling
of][]{2010MNRAS.405.1690D}, and it has been suggested that the small scatter in the SFMS can be understood by applying the central limit theorem to star-forming galaxies \citep{2014arXiv1406.5191K}. Observations typically reveal a scatter of 0.21-0.39 dex (this reflects the range of values given in Table 9 in \citealt{2013ApJ...770...57B}), but
the measured scatter will of course depend on the exact selection
criteria for the star-forming galaxies which form the main sequence
\citep[e.g.][]{2012ApJ...754L..29W} as well as on the uncertainties in
SFR indicators.

\begin{figure}
\centering
\includegraphics[width = 0.45\textwidth]{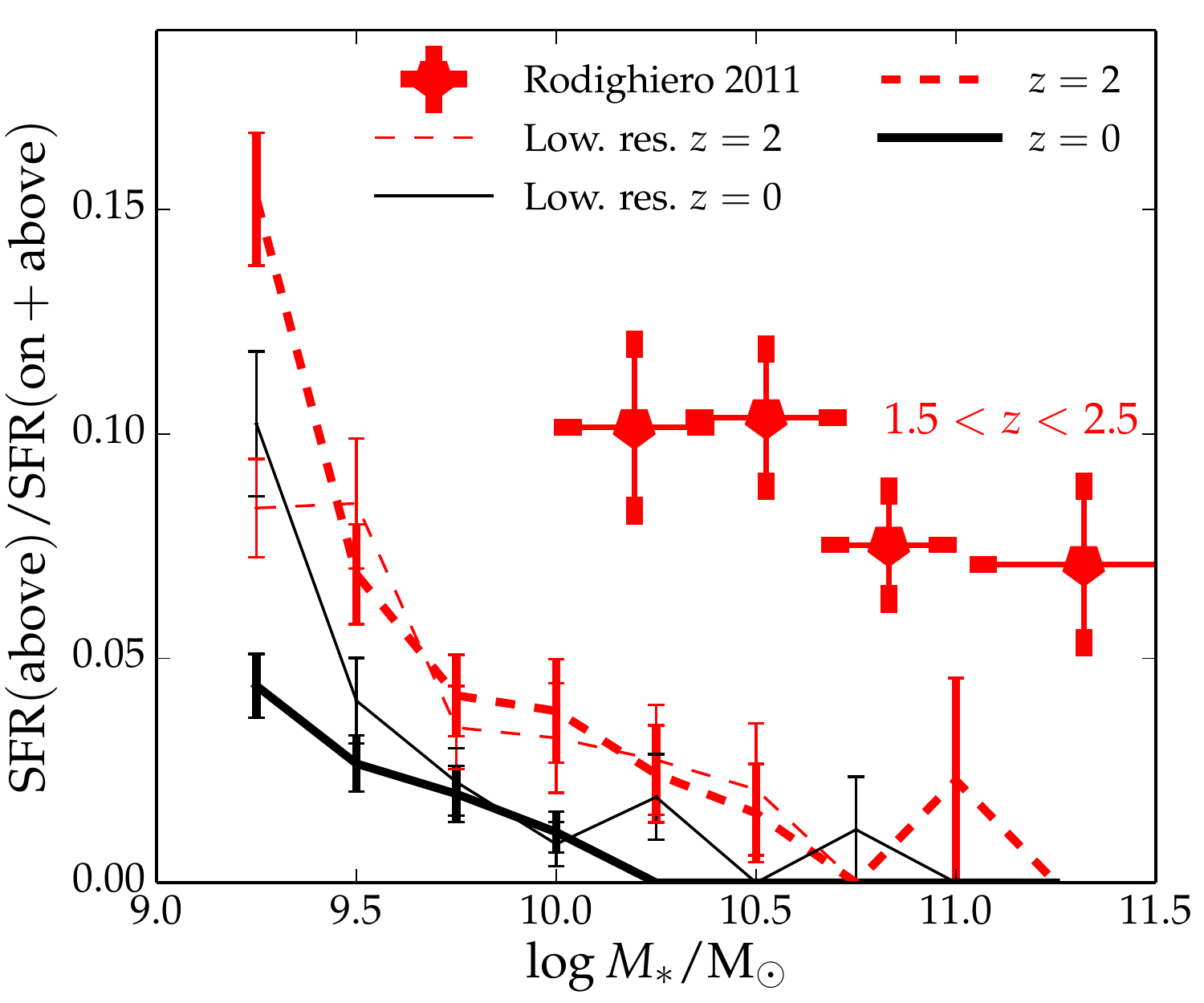}
\caption{The fraction of stellar mass formed by galaxies $2.5\sigma$ or higher above
the SFMS at $z=0$ and $z=2$ (solid and dashed, respectively). The thick and thin lines show the high- and low-resolution
Illustris runs, respectively. The error bars indicate the contribution of Poisson noise. The
simulated galaxy distribution is compared to the observational measurement at $1.5<z<2.5$ from \citet{2011ApJ...739L..40R}.}
\label{Fig_StarburstFraction}
\end{figure}

\begin{figure}
\centering
\includegraphics[width = 0.47\textwidth]{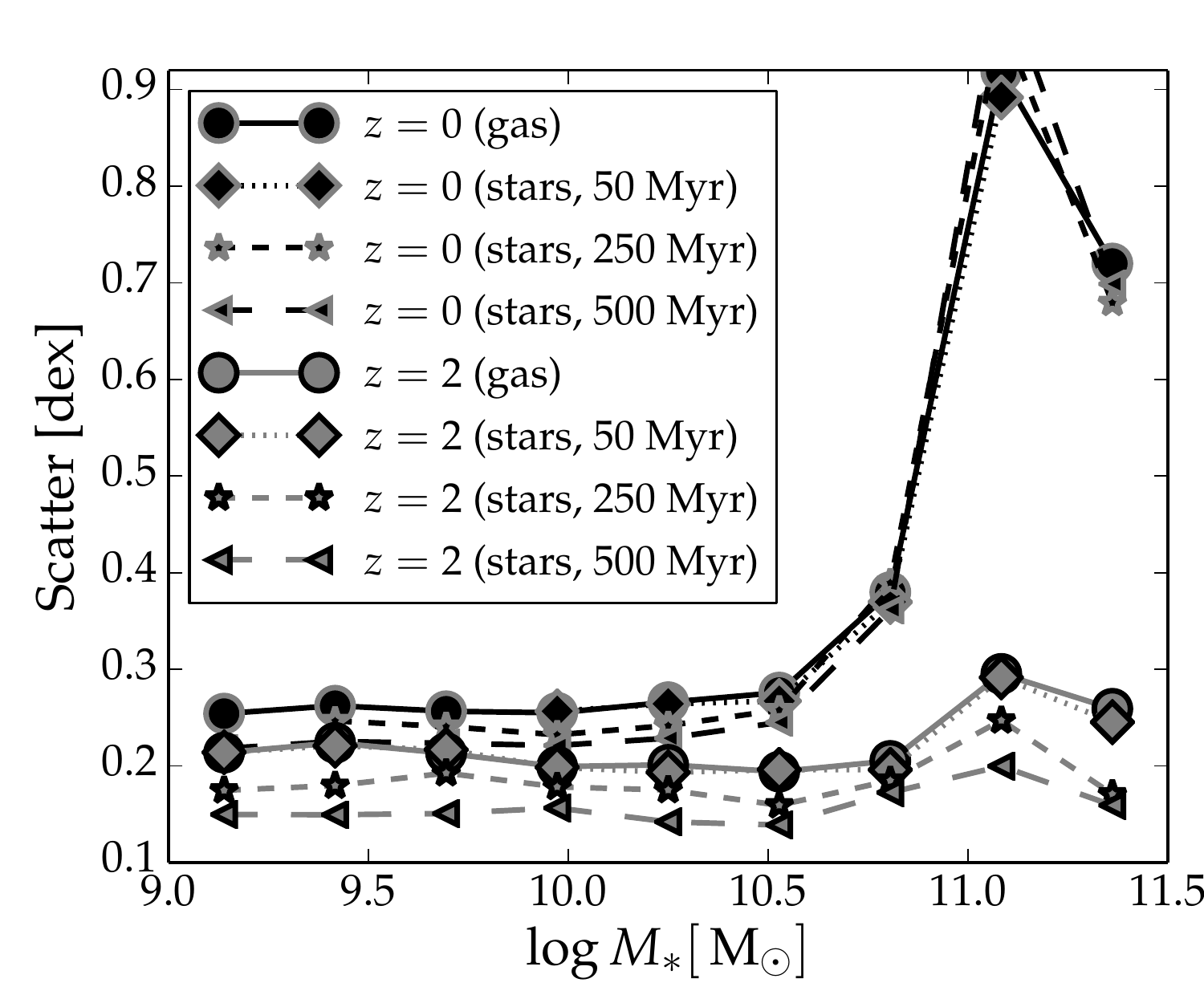}
\caption{The main sequence scatter at $z=0$ and $z=2$ with the SFR
  measured in four different ways: from the instantaneous gas
  properties, and from the initial mass of the stars formed in the
  last 50, 250 and 500 Myr in the galaxies. When estimating the SFR
  from the stars formed in the last 500 Myr, the scatter is
  lower than in the other cases.}
\label{Fig83_MSScatter_SFRIndicator}
\end{figure}

To measure the scatter in the SFMS, we perform a Gaussian fit to the distributions
of SFRs in different stellar mass bins. This method is similar to what is used in
\citet{2011ApJ...739L..40R}.
In Figure~\ref{Fig82_MSScatter}, we show the scatter in the Illustris
galaxies' SFR values as function of stellar mass for four different
redshifts. Below $10^{10.5} \msun$, the scatter is constant at
$0.2-0.3$ dex at each redshift, which is in excellent agreement with
observational constraints. At the high-mass end ($M_*>10^{10.5}
\msun$), the scatter deviates from the value at lower masses because
the galaxy main sequence is ill-defined at these high masses in
Illustris (this is also seen in Figure~\ref{MainSequenceEvolution}).

\subsection{Star formation above the main sequence relation}\label{LesserRoleStarbursts}

After having examined the behaviour of the main sequence relation and
its scatter in the Illustris simulation, we will now look at starburst
galaxies that lie significantly above the main sequence relation.  In
Figure~\ref{Fig_StarburstFraction} the fraction of stars formed in
galaxies on and above the main sequence is computed at different
redshifts. We define a galaxy to be on the main sequence relation if
the SFR is within $2.5\sigma$ of the SFMS relation, which is the same criterion
used in \citet{2011ApJ...739L..40R}. For $M_*>10^{10}
\msun$ the fraction of stellar mass formed in galaxies above the SFMS relation is lower than the
observational result reported by \citet{2011ApJ...739L..40R} at
$1.5<z<2.5$. We also note that the fraction of star formation that occurs above the main sequence is consistent in the high- and
low-resolution Illustris runs for $M_*>10^{9.5}\msun$. In Section~\ref{LowStarburstFraction} we will further discuss the paucity of starbursts in Illustris.

\subsection{The choice of SFR indicator} \label{SFR_indicator}

The SFR of a galaxy can be inferred using different diagnostics, such
as the H$\alpha$ luminosity, the ultraviolet luminosity or the total
infrared luminosity \citep{2012ARA&A..50..531K}. Different diagnostics
are sensitive to the past SFR in a galaxy smoothed over different
timescales, so if a galaxy has a very rapidly changing SFR, different
indicators will yield different SFR values. The timescale for which
different indicators are sensitive can vary from tens to hundreds of
megayears. A situation in which one has to be particularly careful
to rely on such averaging is in the post-starburst phase of merging
galaxies, where simulations show that the SFR inferred from the total
infrared luminosity overestimates the actual SFR of a galaxy by as
much as two orders of magnitudes \citep{2014arXiv1402.0006H}.

\begin{figure*}
\centering
\includegraphics[width = 0.85 \textwidth]{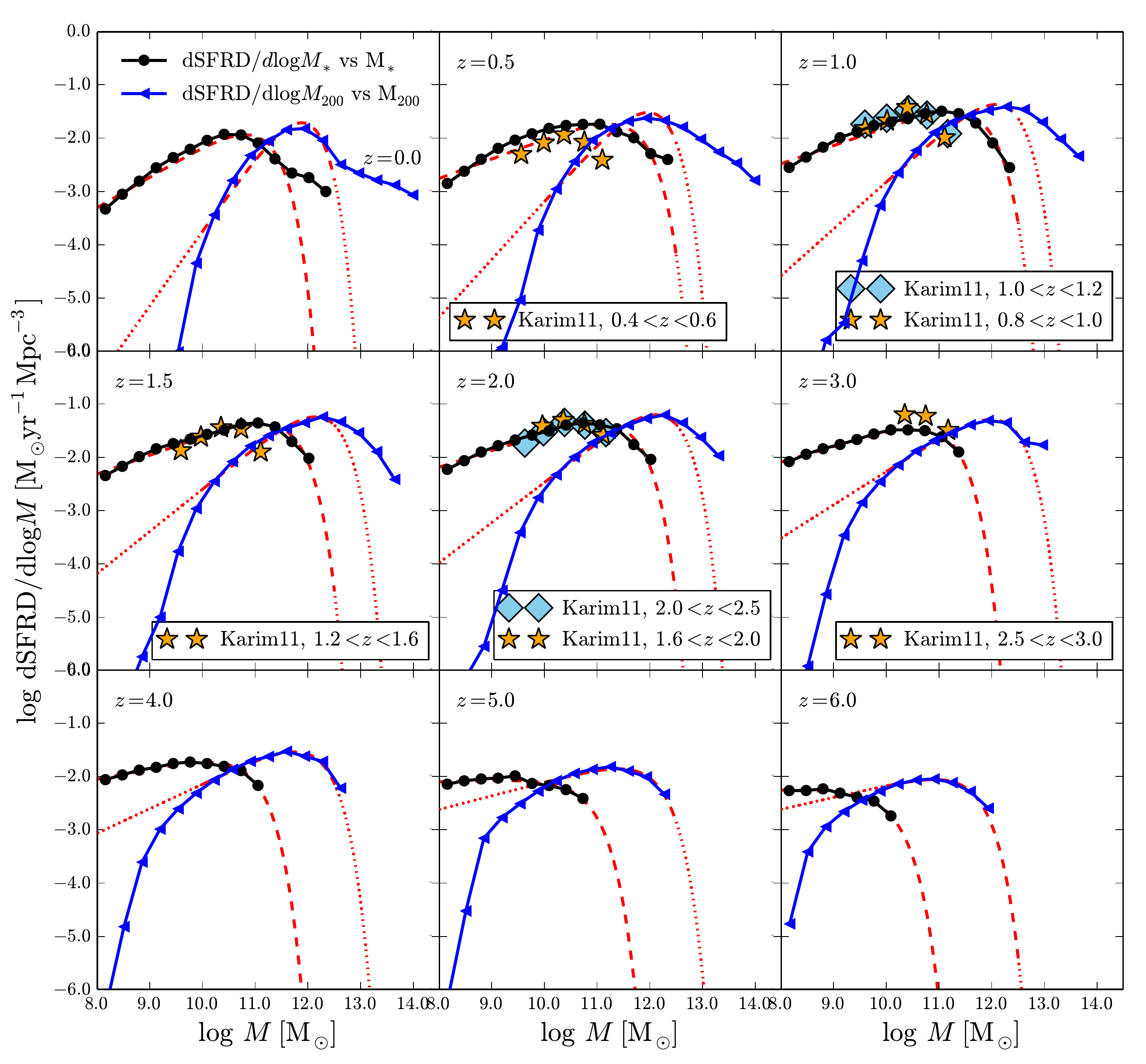}
\caption{The contribution of halos with different stellar masses
  (black circles) and halo masses (blue triangles) to the
  total SFRD in the Universe at different redshifts. The red dashed
  lines are Schechter functions of the form of
  Eq.~\eqref{schechter_formula} fit to the simulation data (outside the fit range the extrapolated Schechter functions are dotted). The distributions evolve with redshift as halos and galaxies grow. This is also seen in Figure~\ref{Fig_56_SchechterFits}, where evolution of the peak mass is examined. Observations from \citet{2011ApJ...730...61K} are shown as stars or squares.}
\label{Fig_55_dSFRDdMstar}
\end{figure*}

In Figure~\ref{Fig83_MSScatter_SFRIndicator} the scatter in the main
sequence is shown for four different definitions of the SFR. First,
the SFR is calculated from the instantaneous gas properties of
galaxies, which we regard as the true SFR of the galaxies. We
additionally calculate the mean SFR from the mass of stars formed in
the last 50, 250 and 500~Myr in a galaxy. The scatter in the main
sequence is essentially the same in the cases where the SFR is
calculated from the gas or the stars formed during the last 50
Myr. For the case where the SFR is calculated from the mass of stars
formed in the last 250 Myr, the scatter in the main sequence decreases
by 0.03 dex at both $z=0$ and $z=2$, and when averaged over 500 Myr,
the scatter decreases by 0.05 dex. It is not surprising that the
scatter declines when increasing the time over which the SFR is
averaged, since the SFHs of galaxies are more similar
when variability on a timescale smaller than e.g. 500 Myr is smoothed
out.

For actively star-forming galaxies, most of the widely used SFR indicators
are sensitive to timescales
smaller than 200 Myr \citep{2012ARA&A..50..531K}. Thus, we
conclude that the timescale over which the SFR indicator is
sensitive is very unlikely to change the derived scatter in the main
sequence relation for the physics model used in Illustris, for which star
formation is less bursty than in reality. The
systematic offsets between different indicators are likely of much greater
importance.

It is possible that the role of the characteristic timescale of an SFR
indicator will have an impact on the derived SFR for feedback
models other than the one used in Illustris. \citet{2013arXiv1311.2073H} and
\citet{2014arXiv1407.0022G}, for example, present feedback models with
typical variability timescales of 10-100 Myr. In
Section~\ref{Sec:500Sample} we further discuss the characteristic
variability timescales of feedback models.

\section{Halo mass and star formation properties of galaxies}\label{Section:GlobalSFR}

\subsection{Halo and stellar masses of star-forming galaxies}

Closely related to the main sequence of star forming galaxies is the
cosmic comoving SFR density (SFRD). Assuming that all galaxies lie on
the main sequence relation, the SFRD can be calculated as
\begin{align}
\text{SFRD} = \int \text{SFR}_\text{SFMS} (M_*) \times \frac{{\rm
    d}n}{{\rm d}M_*} \, {\rm d}M_*,
\end{align}
where $n$ is the comoving number density of galaxies, $M_*$ is the
stellar mass, and SFR$_\text{SFMS} (M_*)$ is the main sequence
relation. The SFRD is observed to peak at $z\simeq 2$
\citep{1996ApJ...460L...1L, 1998ApJ...498..106M, 2004ApJ...615..209H,
  2006ApJ...651..142H, 2010ApJ...716L.103L, 2012A&A...539A..31C}, with
the physical drivers of the evolution being the build-up of massive
halos and the suppression of star formation by feedback from stellar
winds and AGN \citep{2010MNRAS.402.1536S, 2013MNRAS.436.3031V,
  2014MNRAS.438.1985T}.

Figure~\ref{Fig_55_dSFRDdMstar} shows how galaxies with different
stellar masses and halo masses ($M_{200}$) contribute to the SFRD. This has been
computed by summing the contribution to the SFRD from galaxies in
equally spaced logarithmic mass-bins from $10^8\msun$ to
$10^{14.5}\msun$. Bins containing 7 or fewer galaxies are excluded
from the plot in order to avoid bins with very high Poisson
noise. The plot also shows how Schechter functions of the
form,
\begin{align}
\text{SFRD} (M) \propto \left( \frac{M}{M_{\text{sch}}} \right)^\alpha \times \exp \left( -\frac{M}{M_{\text{sch}}} \right),\label{schechter_formula}
\end{align} 
fit to the measurements, where $M$ is either $M_{200}$ or
the stellar mass in a halo and $M_{\text{sch}}$ and $\alpha$ are free
fit parameters. For the fits in terms of halo mass, we
exclude halos outside the range $10^{10}\msun
<M_{200}<10^{12.5}\msun$, since a Schechter function does not yield a
good fit over the entire range of $M_{200}$-values of the halos. For fits in terms
of stellar mass, all galaxies with $10^{8}\msun
<M_*<10^{12.5}\msun$ are included. The actual distributions measured
from the Illustris simulation are overall quite well described by the
Schechter fits. This is consistent with the observations from
\citet{2011ApJ...730...61K}, where the distributions of dSFRD$/$d$\log
M_*$ are also well fit by a Schechter function for (at least)
$M_*>10^{8}\msun$. When comparing dSFRD$/$d$\log M_*$ in Illustris with the observations from \citet{2011ApJ...730...61K}, some tensions are visible. At $z=0.5$, the normalisation of this function is too large in Illustris, and at $z=3$, it is too low. Deviations are, of course, expected, since dSFRD$/$d$\log M_*$ is closely linked to the SFMS and stellar mass function in Illustris, which both show tensions with observations. At $z=1$, the figure shows good agreement between Illustris and \citet{2011ApJ...730...61K}. This is, however, a coincidence: the star formation rates of galaxies are $\sim0.5$ dex too low in Illustris at this epoch (see Figure~\ref{Fig:SSFR_Redshift}), but the normalisation of the stellar mass function is too large at low masses. Together, these effects yield a dSFRD$/$d$\log
M_*$ in agreement with observations.

\begin{table*}
\begin{tabular}{ccccc} 
\hline\hline 
$M_{200}/\msun$ &  median($M_{200}$)$/\msun$ &  $\min( M_{200} )$/median($M_{200}$) & $\max( M_{200} )$/median($M_{200}$)    &  $N_\text{galaxies}$\\
\hline
$10^{11}$ & $10^{11}$ & 1.00 & 1.00 & 100 \\
$10^{12}$ & $10^{12}$ & 0.97 & 1.03 & 100 \\
$10^{13}$ & $9.2\times 10^{12}$ & 0.77 & 1.42 & 100 \\
\hline\hline
\end{tabular}
  \caption{Three samples with different halos masses, $M_{200}$. Each sample includes the 100 galaxies with masses closest to $10^{11}$, $10^{12}$ and $10^{13} \msun$ at $z=0$.}
\label{table:HaloMassRanges}

\end{table*}

\begin{figure}
\centering
\includegraphics[width = 0.48 \textwidth]{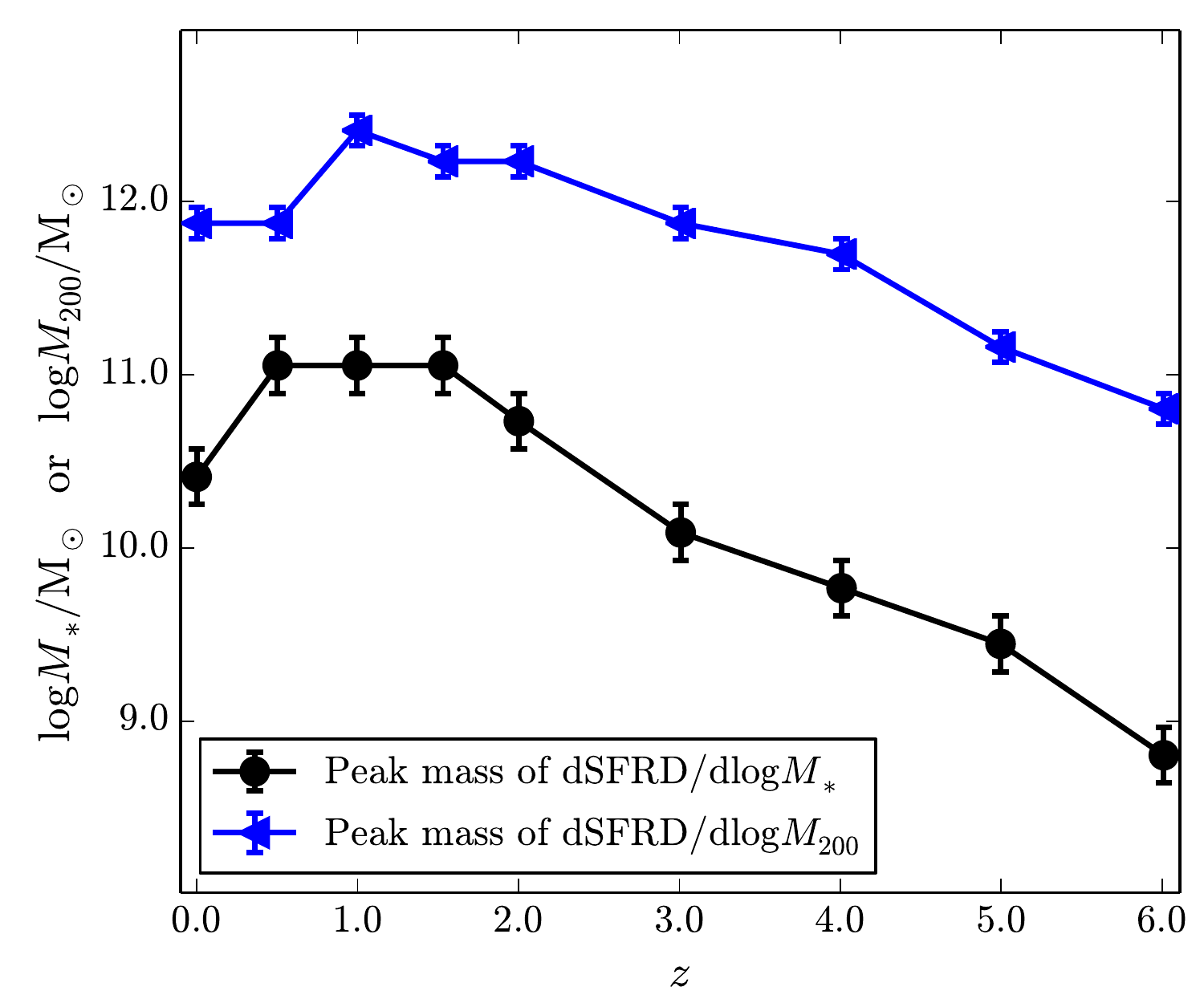}
\caption{The redshift evolution of the peak stellar mass and halo mass for
  dSFRD$/$d$\log M_{200}$ and dSFRD$/$d$\log M_*$, respectively, for
  the distributions in Figure~\ref{Fig_55_dSFRDdMstar}. The peak mass (for both stars and halos) increases from $z=6$ to $z\simeq 1-2$, after which it turns over. The error bars are set by the bin width of the histograms shown in Figure~\ref{Fig_55_dSFRDdMstar}. }
\label{Fig_56_SchechterFits}
\end{figure}

\begin{figure}
\centering
\includegraphics[width = 0.48 \textwidth]{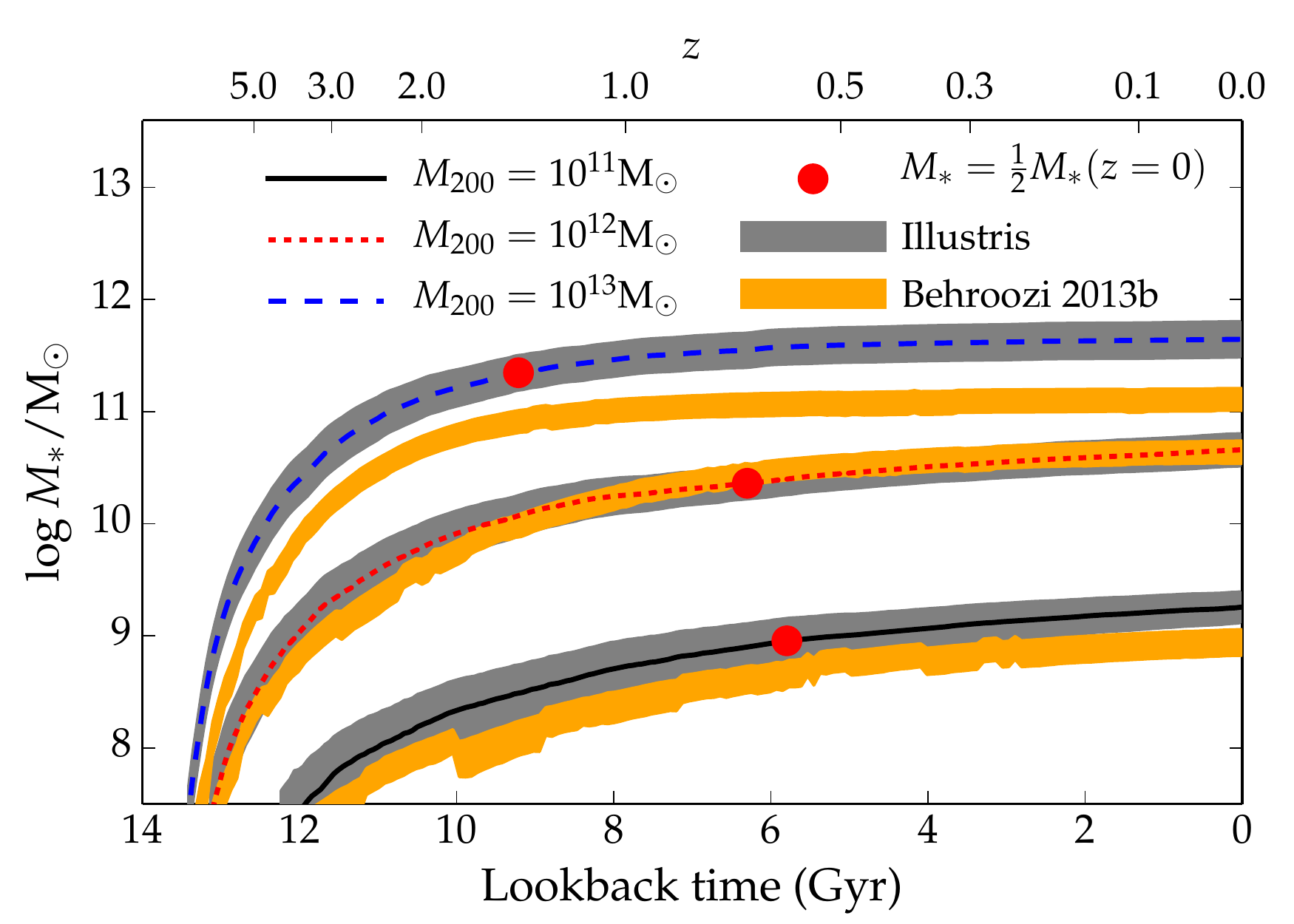}
\caption{The fraction of the current stellar mass formed as function
  of lookback time for galaxies residing in halos with
  $M_{200}$-values of $10^{11} \msun$, $10^{12} \msun$ and $10^{13} \msun$ (see
  details about samples in Table~\ref{table:HaloMassRanges}). The
  grey shaded regions correspond to $1\sigma$ scatter in each mass
  range. The orange regions are the $1\sigma$ confidence intervals
  from the abundance matching of \citet{2013ApJ...770...57B} for
  $M_{200}/\msun=10^{11} \msun$, $10^{12} \msun$ and $10^{13} \msun$.}
\label{Fig_StellarMassGrowthHistories}
\end{figure}

The peak masses of the distributions for dSFRD$/$d$\log M_{200}$ and
dSFRD$/$d$\log M_*$ are plotted at different redshifts in
Figure~\ref{Fig_56_SchechterFits}. At $z\leq 4$ the peaks in the
distributions occur at $10^{11.5}<M_{200}/\msun<10^{12.5}$ and
$10^{10.0}<M_*/\msun<10^{11.0}$. A large contribution of star formation from halos
with $M_{200}=10^{12}\msun$ is, e.g., also seen in the abundance
matching analysis of \citet{2013A&A...557A..66B}. 

At $M_{200}\simeq 10^{12} \msun$, halos are most efficient in
turning their baryons into stars. This is for example evident when
examining $M_*/M_{200}$, which peaks around $M_{200}\simeq
10^{12} \msun$ (this relation is plotted for Illustris
in \citealt{2014arXiv1405.2921V} and \citealt{2014arXiv1405.3749G}). This is because AGN feedback suppresses
the formation of stars in halos above this characteristic mass, and
stellar winds are responsible for suppressing the formation of stars
in lower mass halos. It is therefore not surprising that the peak in
dSFRD$/$d$\log M_*$ is present at $M_{200}\simeq 10^{12}
\msun$ at $z\lesssim 4$. At $z\gtrsim 4$
there is a decline in the typical masses (both stellar and halo
masses) at which stars are formed in Illustris, since halos and
galaxies are less massive at high redshifts.

A feature that is also visible in Figure~\ref{Fig_56_SchechterFits} is that the mass (both for halo and stellar mass) at which most of the star formation occurs declines from $z=1$ to $z=0$. This is consistent with the downsizing scenario, where the most-massive galaxies at $z=0$ have older stellar populations than less-massive galaxies.

\subsection{Build up of stellar components at different halo masses}

To study how halo mass affects the average formation history of stars
we create samples of halos with $M_{200}=10^{11},
10^{12}$ and $10^{13}\msun$. Each sample contains the 100 galaxies
with halo masses closest to the mass that defines the
samples. Table~\ref{table:HaloMassRanges} summarizes the median,
minimum and maximum halo mass in each sample. For each galaxy in each
sample we compute the stellar mass growth history by calculating the
amount of a galaxy's stellar mass at $z=0$ that was formed at
different lookback times. For this analysis we include all stars within a halo at $z=0$, including the stars residing in subhalos.

Figure~\ref{Fig_StellarMassGrowthHistories} shows the median stellar
growth history for each of the three samples. The galaxies ending up
in the most-massive halos considered ($\sim10^{13}\msun$) form
their stars much earlier than the galaxies ending up in the least
massive halos. This is most easily seen by comparing the times at
which half of the stellar mass present at $z=0$ is formed (shown as
red circles in the figure). It is perhaps not surprising that very
massive halos form their stars earlier than less massive ones, since
halos of $10^{12}\msun$ contribute most to the global SFR in the
Universe, as shown above. However, this trend runs counter to the
formation time of the dark matter halos themselves, where the most
massive halos form latest as a result of hierarchical structure
growth.

We compare Illustris to the results from \citet{2013ApJ...770...57B},
where the stellar growth history is derived from the stellar mass --
halo mass relation ($M_* (M_\text{h},z)$), which is determined by
comparing dark matter merger trees with observations of the cosmic
SFR, the specific SFR vs. $M_*$ and stellar mass functions. For
$10^{12}\msun$ halos, we find excellent agreement between Illustris
and \citet{2013ApJ...770...57B}. At $10^{11}\msun$ and
$10^{13}\msun$, Illustris predicts stellar components roughly 0.4
dex more massive than \citet{2013ApJ...770...57B}. This overproduction
of stellar mass in halos of mass $\sim10^{11}\msun$ and
$\sim10^{13}\msun$ is essentially independent of redshift.
\citet{2014arXiv1405.3749G} and \citet{2014arXiv1405.2921V} found similar deviations between the Illustris simulation and abundance matching results when analysing the high- and low-mass ends of the stellar mass function. \citet{2014arXiv1405.3749G} regard these tensions as significant and suggest that more realistic feedback models could potentially help suppress the formation of too-massive stellar components in low- and high-mass galaxies.

\begin{table*}
\begin{tabular}{ccccc} 
\hline\hline 
$M_*/\msun$ &  median($M_*$)$/\msun$ &  $\min( M_* )$/median($M_*$) & $\max( M_* )$/median($M_*$)    &  $N_\text{galaxies}$\\
\hline
$10^9$ & $1.0\times 10^9$ & 0.99 & 1.02 & 500 \\
$10^{10}$ & $1.0\times 10^{10}$ & 0.95 & 1.05 & 500 \\
$10^{11}$ & $9.4\times 10^{10}$ & 0.83 & 1.30 & 500 \\
$>1.57\times 10^{11}$ (most massive) & $2.6\times 10^{11}$ & 0.60 & 12.70 & 500 \\
\hline\hline
\end{tabular}
\caption{Four different samples of galaxies in different
  stellar-mass-ranges. 
  The first three samples include the 500 galaxies with stellar masses ($M_*$) closest to $10^9$, $10^{10}$ and $10^{11} \msun$ at $z=0$. The fourth range includes the 500 galaxies with the largest stellar masses at $z=0$. The median, minimum and maximum stellar mass for each sample is also listed.}
\label{table:MassRanges}
\end{table*}

\section{Star formation histories of galaxies}\label{Sec:500Sample}

\subsection{Outliers from the average star formation histories}

An alternative to studying statistical properties of galaxies with
different stellar masses is to analyse their individual star formation
histories in more detail. To compute the star formation history (SFH)
we select the stars ending up in a galaxy at $z=0$, and then create a
histogram of the initial mass of stars formed in 100 equally spaced
bins in terms of lookback time. We include all stars that end up in
the main stellar component of a galaxy (i.e. we exclude satellites) when
calculating the SFH in this way. We do not
distinguish between stars formed in-situ or ex-situ. We study galaxies
in four different stellar mass ranges at $z=0$. Three ranges are
centered around $M_*=10^9 \msun$, $M_*=10^{10} \msun$ and
$M_*=10^{11} \msun$, with the widths of the different ranges chosen
such that they include 500 galaxies each. We additionally create a
range of the 500 galaxies with the most-massive stellar
components. Basic properties of the different samples are summarised
in Table~\ref{table:MassRanges}.

\begin{figure}
\centering
\includegraphics[width = 0.48 \textwidth]{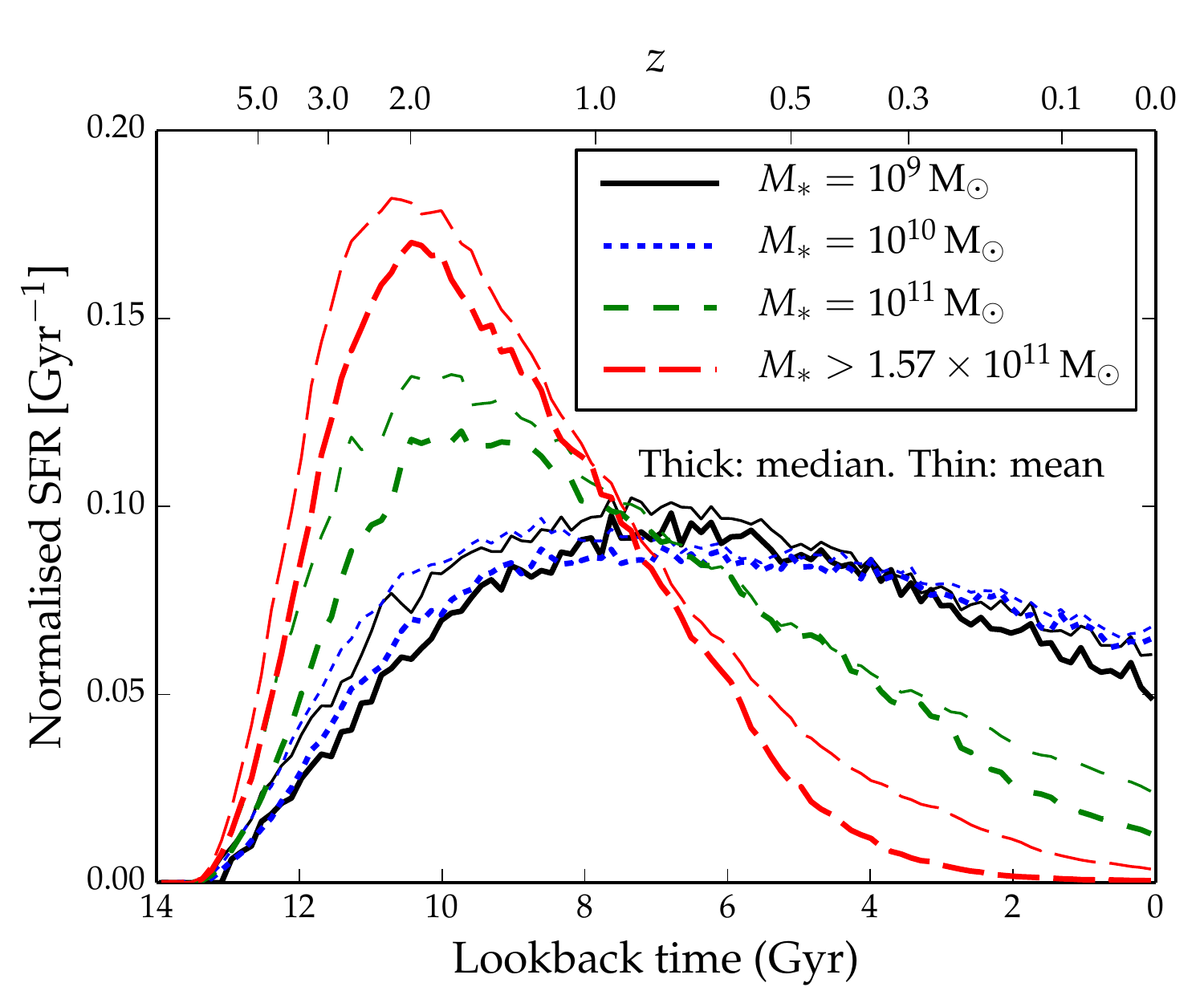}
\caption{Mean and median (thin and thick lines, respectively) star
  formation histories of galaxies in the four different stellar mass
  ranges (see Table~\ref{table:MassRanges}). The star formation rates
  are normalised so that $\int \text{SFR} (t) {\rm d}t = 1$.}
\label{SFR_histories}
\end{figure}

The mean and median star formation histories of the galaxies in the
different mass-ranges are shown in Figure~\ref{SFR_histories}. For the
$10^{11}\msun$ range and the 500 most-massive galaxies the SFH peaks
at $z\simeq 2$, and decreases at later epochs. These trends are
qualitatively in good agreement with other studies of star formation
histories of galaxies \citep{2013ApJ...770...57B,
  2014arXiv1404.0402S}. For the $10^{9}\msun$ and $10^{10}\msun$
mass ranges, the peak is significantly broader, and it occurs at $z\simeq
1$. In general, the mean SFR is 10-20\% larger than the median value,
since the mean is more sensitive to extreme outliers with high SFRs.

\begin{figure*}
\centering
\includegraphics[width = 0.95 \textwidth]{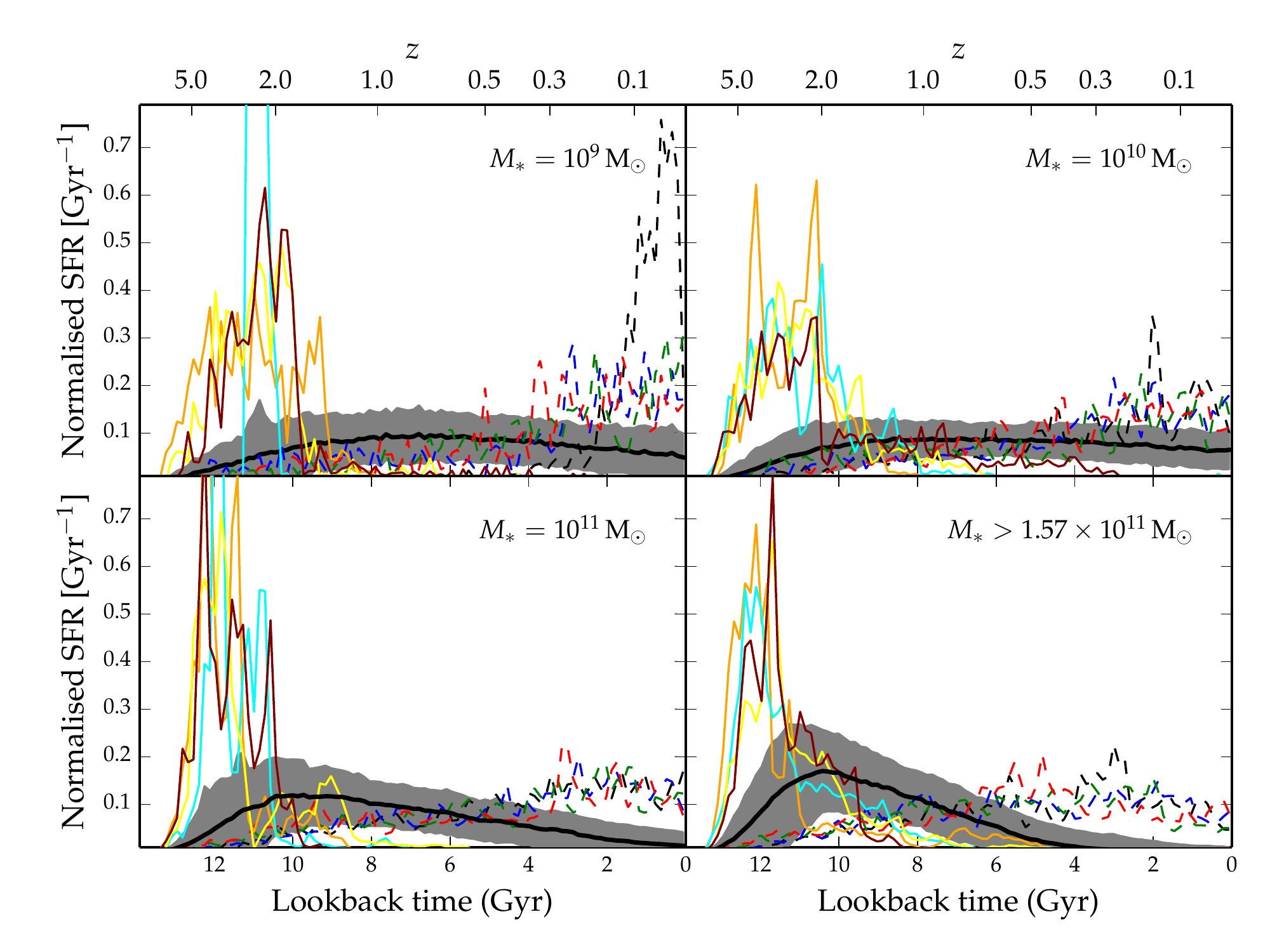}
\caption{For each $M_*$-range, we have selected the four galaxies that
  form 50\% of their stars earliest or latest (solid and dashed thin lines, respectively). The median profiles for each range are shown by the thick black lines, and the 68\% confidence intervals are the shaded regions. The normalisation convention is $\int \text{SFR} (t) {\rm
    d}t = 1 $.}
\label{SFR_histories_outliers}
\end{figure*}

Different galaxies experience a variety of gas accretion and merger
histories, and they are therefore expected to exhibit diverse SFHs. We
illustrate this in Figure~\ref{SFR_histories_outliers}, where we show
for each mass range the SFR of the galaxies that form 50\% of their
stellar mass earliest or latest, compared to the average history of
the corresponding sample. In all mass ranges, it is possible to find 
SFHs with early or late star formation. The galaxies with
decidedly early star formation histories exhibit a similarly bursty
epoch at a lookback time of $\simeq 10-13$ Gyr, and the galaxies with
late star formation histories display a prominent star forming mode
contributing at $z\lesssim 0.5$ and hardly any high-redshift star
formation.

\begin{figure*}
\begin{minipage}[b]{0.45\linewidth}
\centering
\includegraphics[width = \textwidth]{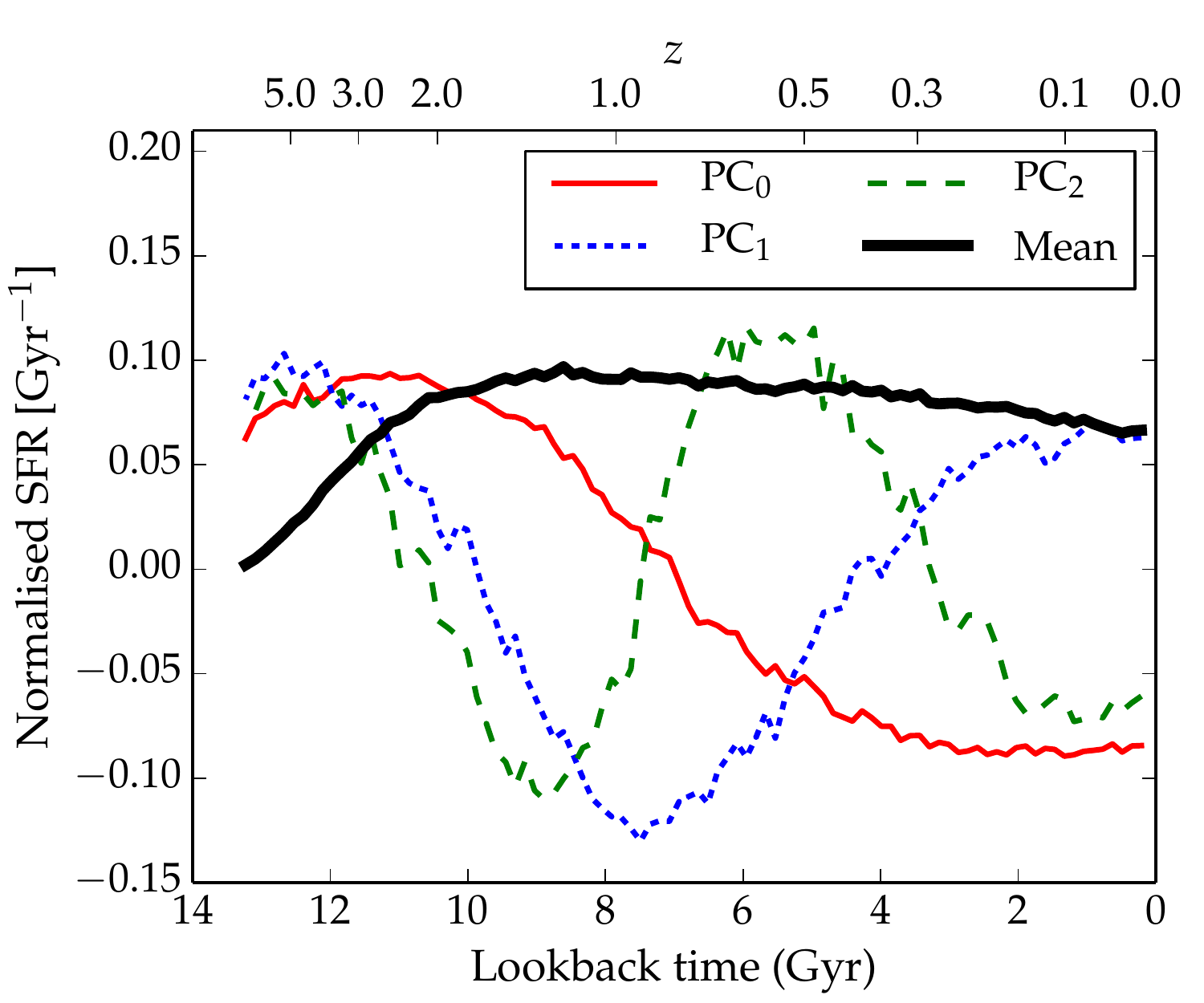}
\caption{The mean SFR for the sample with $M_* = 10^{10} \msun$
  (from Table~\ref{table:MassRanges}) together with the first three
  principal components (PC$_i$ for $i=0,1,2$) describing the
  scatter around the mean. The normalisations of the principal
  components are arbitrary.}
\label{SFR_PCA_Intro}
\end{minipage}
\hspace{0.5cm}
\begin{minipage}[b]{0.45\linewidth}
\centering
\includegraphics[width =  \textwidth]{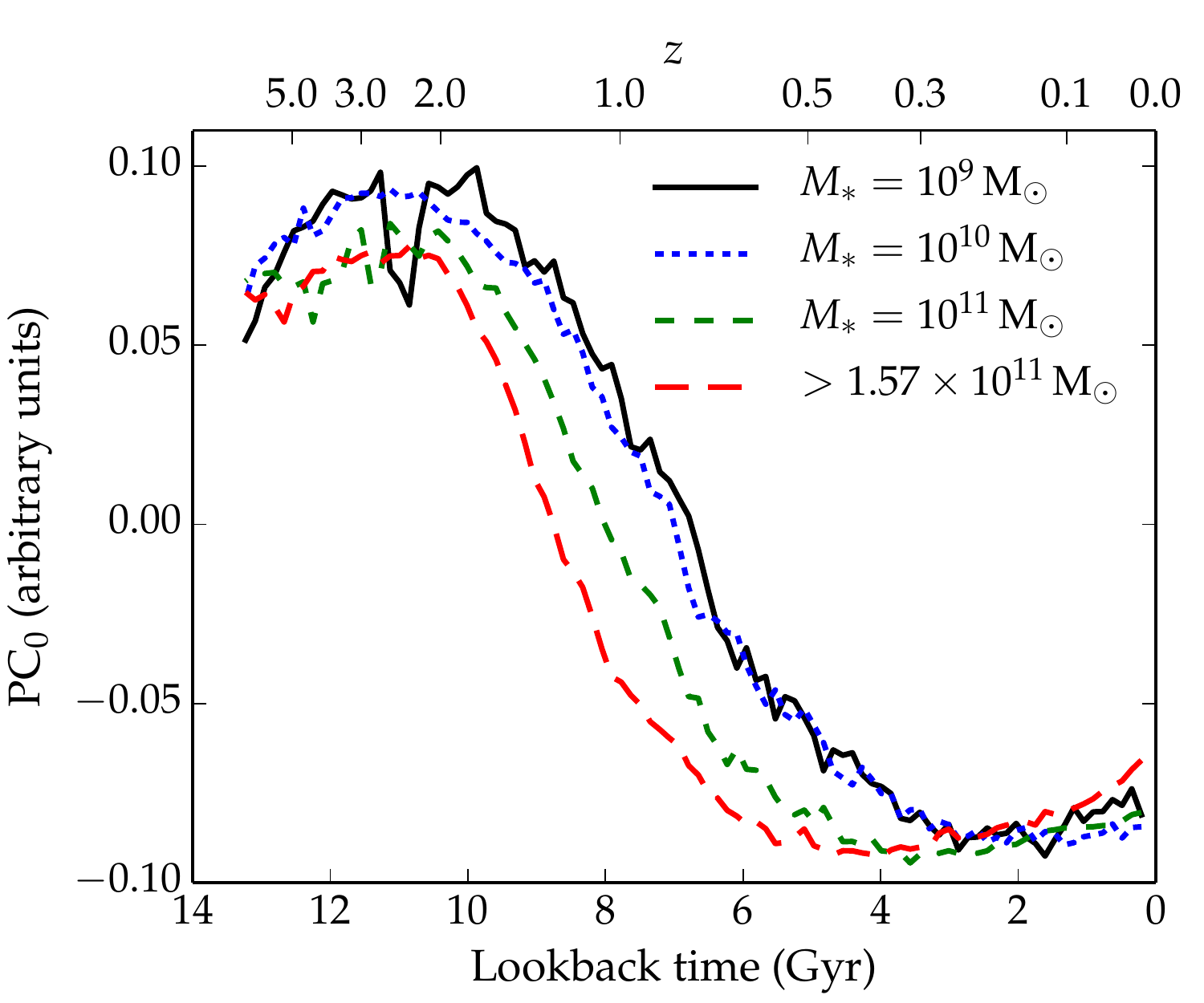}
\caption{The zeroth principal component for the four different mass
  samples in Table~\ref{table:MassRanges}. In all mass samples the
  PC$_0$ component is of similar shape and shows a positive
  contribution at a lookback time of 11 Gyr, and a negative
  contribution at lookback times smaller than $\simeq 7$ Gyr.}
\label{SFR_PCA_All4Bins}
\end{minipage}
\end{figure*}

\subsection{Modes of star formation in a principal component analysis}\label{PCA-subsection}

The relative similarity seen in Fig.~\ref{SFR_histories_outliers} of
the galaxies in the tails of the star formation history distribution,
forming their stars extremely late or early, suggests that a more
systematic study of the star formation modes in the Illustris galaxies
is worthwhile. As a tool for statistically characterizing the star
formation histories we have adopted a \emph{principal component
  analysis} \citep[PCA; inspired by][]{2014arXiv1406.2967C} where the
SFH of a galaxy is seen as a vector in a $N$-dimensional space, where
$N$ is the number of bins used to characterize the SFH of a galaxy (we
use $N=100$). A key quantity in a PCA analysis is the average SFH of
the galaxies in the sample,
\begin{align}
\langle \text{SFR}\rangle (t_i) = \frac{1}{N_\text{galaxies}} \sum_{j=0}^{N_\text{galaxies}-1} \text{SFR}_j(t_i),
\end{align}
for $i=0,\ldots ,N-1$. We furthermore define the scatter matrix,
\begin{align}
C_{mn} =\sum_{j=0}^{N_\text{galaxies}-1} \left[  \text{SFR}_j(t_m)- \langle \text{SFR} \rangle (t_m)    \right]
\left[  \text{SFR}_j(t_n)- \langle \text{SFR} \rangle (t_n)    \right],
\end{align}
Of special interest are the eigenvectors of $C_{mn}$, which
are called the \emph{principal components} (PC$_i$, for $i=0,\ldots,
N-1$) and describe the deviations between the SFH of individual
galaxies and the mean SFH of a sample. Principal components
diagonalize the scatter matrix, implying that the scatter between
different PC$_i$'s is uncorrelated. Conventionally the PCs are ordered
in descending order by the contribution they make to the total
variance.  Often, the first components account for much of the
variance, and the corresponding eigenvectors can be interpreted in
physical terms. In our case they can serve the purpose of
characterizing the most important `modes' of the SFH.

In terms of the principal components, the SFH of a galaxy can be
written as
\begin{align}
\text{SFR} = \langle\text{SFR}\rangle + \sum_{i=0}^{N-1} q_i \times \text{PC}_i,
\end{align}
where $q_i$ is the coefficient determining the strength of the
contribution of PC$_i$ for a specific galaxy. Since we have binned our
SFHs in $N=100$ bins we formally get 100 principal components, and for
each of the 500 galaxies in each sample we obtain 100 coefficients
$q_i$. The limited size of our galaxy set means that only the leading
PC components are robust against the noise.

\begin{figure*}
\centering
\includegraphics[width = 0.95 \textwidth]{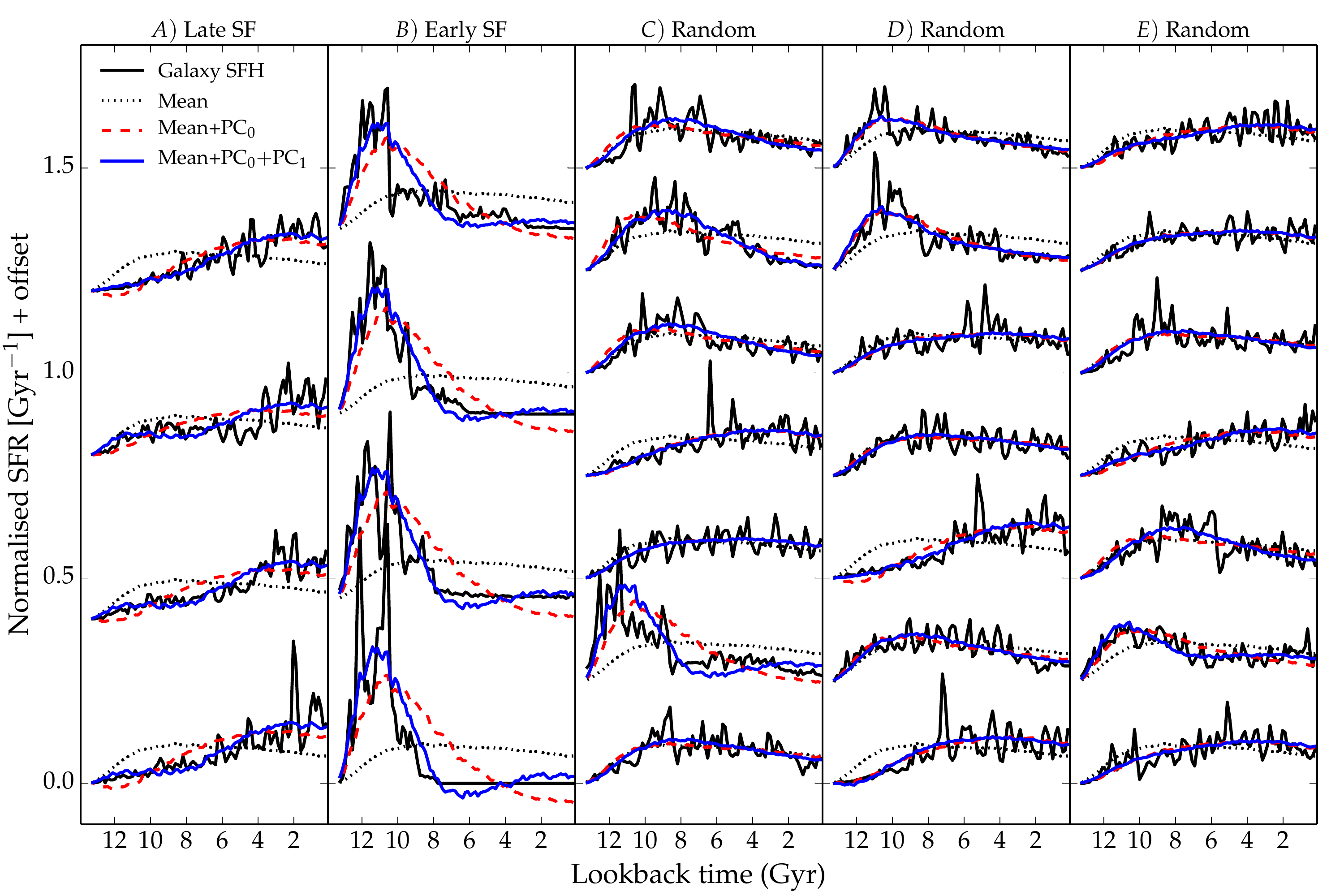}
\caption{Star formation histories of individual galaxies (black lines)
  and the mean profile for the sample of galaxies with
  $M_*=10^{10}\msun$ (grey dashed lines). Also shown are profiles
  including the contribution from the zeroth principal component (red
  dashed line) and components 0 and 1 (blue line). Panels $A$ and $B$
  show 8 galaxies that form their stars late and early, respectively
  (the same profiles are shown in
  Figure~\ref{SFR_histories_outliers}). Panels $C$, $D$ and $E$ show 21
  randomly selected galaxies from the sample with
  $M_*=10^{10}\msun$.}
\label{Fig88_PCA_Histories}
\end{figure*}

Figure~\ref{SFR_PCA_Intro} shows the mean SFH for the $M_*=10^{10}
\msun$ sample together with the three leading modes, PC$_0$, PC$_1$
and PC$_2$. The PC$_0$-mode accounts for galaxies forming stars early
or late, depending on whether the coefficient $q_0$ is positive or
negative. PC$_1$ and PC$_2$ cross the zero-point two and three times,
respectively, and they are therefore determining the more detailed
evolution of the SFH. In general, PC$_i$ crosses the zero-point $i+1$
times, and therefore is associated with a physical timescale of order
$t_\text{H} / (i+1)$, where $t_\text{H}=13.7$ Gyr is the Hubble time.

For all the four mass samples from Table~\ref{table:MassRanges}, the
PC$_0$ is plotted in Figure~\ref{SFR_PCA_All4Bins}. They all appear
similar and have the feature that a positive $q_0$-value describes a
galaxy forming stars earlier than the mean of the sample. The only
remarkable difference between the PC$_0$ eigenvectors from the
different samples is that the lookback time where the change from
positive to negative contributions occurs shifts to higher values for
the more massive galaxies. The trend that massive galaxies have earlier bursty epochs than less massive galaxies is therefore both reflected by the mean star formation history and a 0th principal component that peaks at higher redshift.

Figure~\ref{Fig88_PCA_Histories} compares the actual SFHs for a
selection of galaxies to the mean SFH of the $M_*=10^{10} \msun$
sample, and the SFHs reconstructed by including only PC$_0$ and by
including both PC$_0$ and PC$_1$. By including PC$_0$ it is possible
to capture whether a galaxy forms its stars early or late. Including
the PC$_1$ mode as well gives of course a better fit to the actual SFH
of the galaxies. Components 0 and 1 account for 33\% and 10\% of the
total scatter in the sample. These percentages indicate that while a
given galaxy's SFH can already be described reasonably well by just
PC$_0$ and PC$_1$, variability on timescales much shorter than
described by them needs to be considered for an accurate description.

We note that it could potentially be useful to construct a family of SFHs
based on a combination of the mean SFH for the Illustris galaxies and one
or more of the principal components from the PCA analysis. An attractive
property of a model based on the mean SFH and the leading principal
component would be that it could describe a realistic SFH of a galaxy
with only one free parameter (the relative contribution of the mean SFH
and the leading principal component). Such a model would be more physically
motivated than e.g. single-burst models or $\tau$-models, which are often
used when fitting spectral energy distributions of galaxies. In future work
we will construct such a model and show how it can be used for fitting
spectral energy distributions. Such an approach could yield significant
advantages because the accuracy and robustness of spectral energy distribution
modelling can be very sensitive to the assumed SFH
\citep[e.g.][]{Michalowski2012,Michalowski2014,HS2014,Torrey2014}.

\subsection{Assessing the SFH variability timescale}

\begin{figure}
\centering
\includegraphics[width = 0.48 \textwidth]{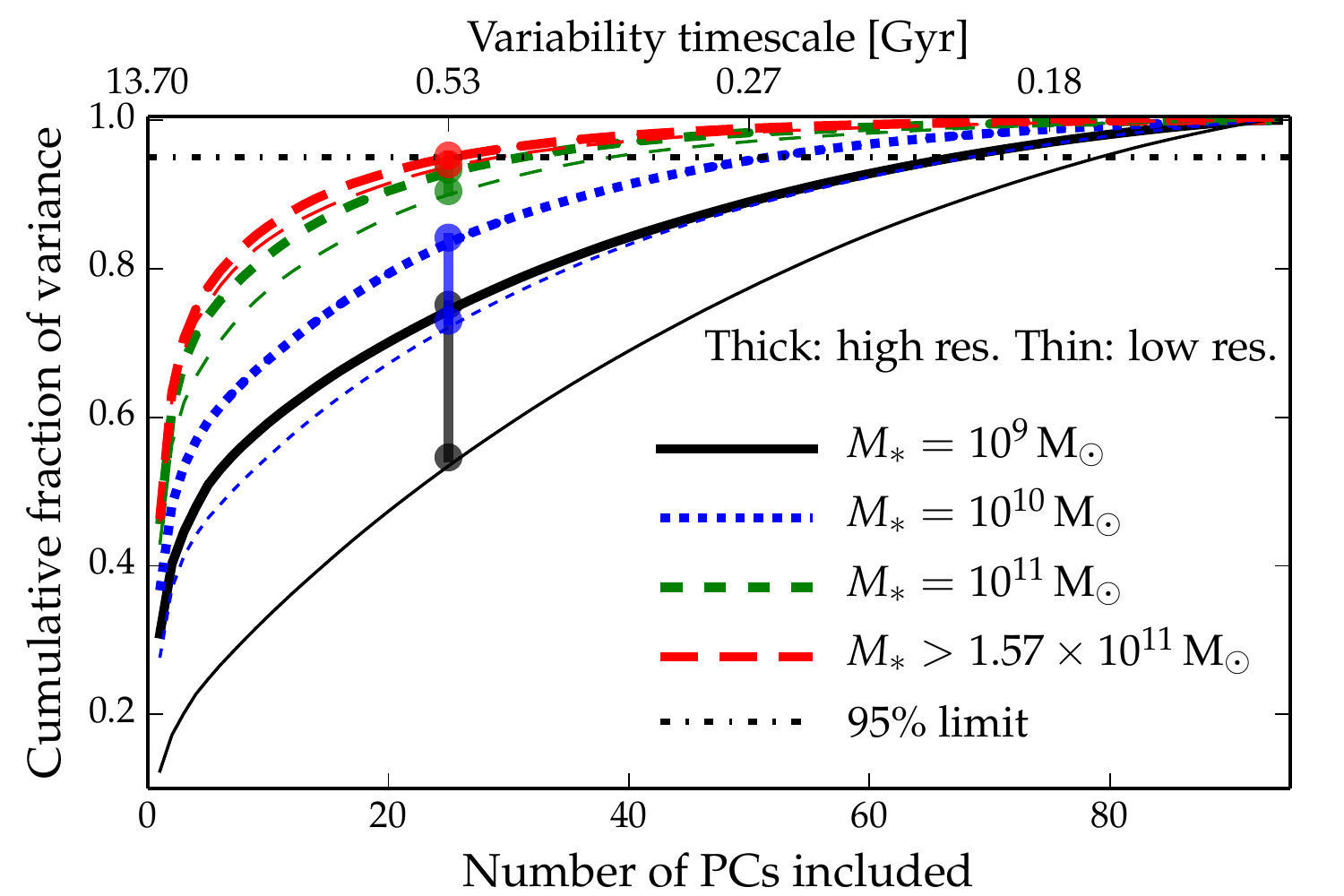}
\caption{The cumulative fraction of the total variance in the star
  formation histories accounted for as a function of the number of
  principal components that are included. The thick lines show the full resolution run and the thin lines show the low resolution run. The vertical lines (connected by circles) measure how well-converged the variability timescale is for the different mass-ranges.}
\label{Fig89_PCA_FracQ}
\end{figure}

As each principal component has an associated timescale, the PCA
analysis may also be used to characterize variability of the star
formation histories, which in turn is influenced by the adopted
physical model for feedback processes and the ISM. In
Figure~\ref{Fig89_PCA_FracQ}, we show the cumulative fraction of the
variance accounted for as a function of the number of principal
components included. The two most-massive samples behave similarly for the
low and high resolution run. For the two low mass bins the difference between
the low and high resolution run is larger. To quantitatively measure
how well-converged the cumulative fractions in Figure~\ref{Fig89_PCA_FracQ}
are, we use the difference in the fraction of the scatter accounted for by
the 25 first components in the low and high resolution run. For the various
bins the differences are 21\%, 11\%, 3\% and 1\% (from lowest to highest mass).
The two most-massive ranges are therefore well-converged, unlike
the two least-massive ranges, where some of the scatter comes from Poisson noise.
This is also consistent with Figure~\ref{SFR_histories}, where the mean and
median star formation history of the low-massive range seem to be more affected
by noise the most-massive range.

The timescale at which 95\% of the scatter is accounted for is 500 Myr
in the two most-massive bins, and for the other mass-ranges this timescale
is smaller, which might be a consequence of the contribution of Poisson noise
being more important in the low-massive galaxies, where there are fewer star
particles per time bin. Based on the variability timescale of the most-massive range,
the characteristic timescale for fluctuations in the galactic star formation for the
ISM model adopted in Illustris has characteristic timescale of 500 Myr. In
Section~\ref{SFR_indicator} we saw that the scatter in the SFMS decreased
significantly (by 0.05 dex), when the SFH was smoothed on this timescale, and
that the decrease in the scatter was more moderate when smoothing over shorter
timescales.

We note that there exist other simulation feedback models which show
variability timescales an order of magnitude lower than
Illustris. \citet{2013arXiv1311.2073H}, for example, include a
treatment of radiative feedback from young stars and more localized
supernova feedback in giant molecular clouds. Another model with a
short variability timescale is \citet{2014arXiv1407.0022G}, where this
is achieved through a high-star formation threshold and delayed
radiative cooling. Interestingly, variability itself could be used as
an important constraint of feedback models, although the fact that
highly variable feedback models will induce differences between SFR
indicators may make this complicated in practice.

\begin{figure*}
\centering
\includegraphics[width = 0.8 \textwidth]{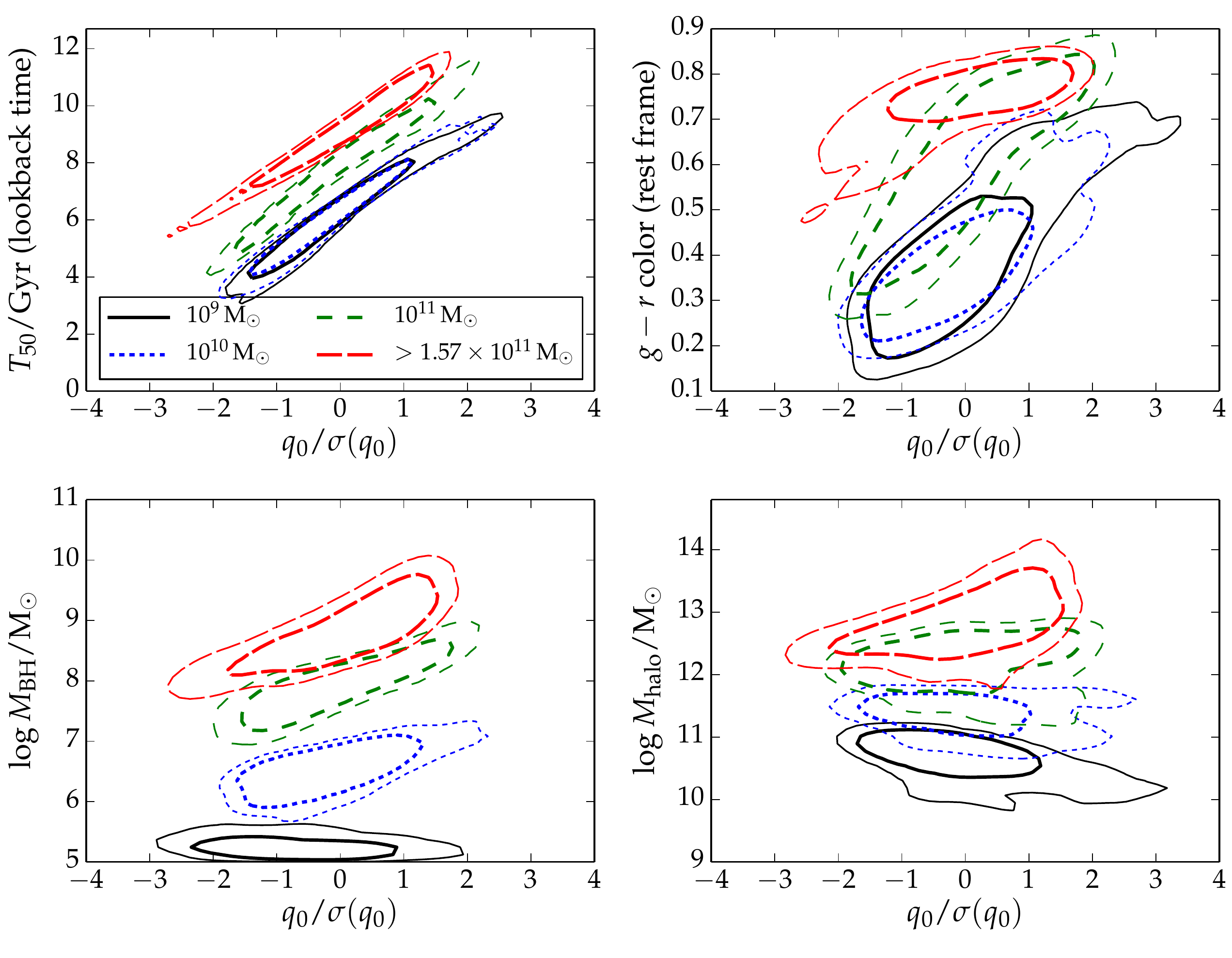}
\caption{The dependence of various galaxy properties on the
  coefficient describing the contribution from the zeroth principal
  component ($q_0$). The thick and thin contours show 68\% and 90\% confidence intervals, respectively, for the galaxies in the samples with stellar masses
  of $M_*/\msun = 10^{9},10^{10},10^{11}$, as well as for the 500
  most-massive galaxies. The \emph{upper left panel} shows that the lookback
  time at which 50\% of the stars in a galaxy is formed ($T_{50}$)
  correlates strongly with $q_0$. There also exists correlations with
  the $g-r$ colour (\emph{upper right panel}), and the black hole mass at
  $z=0$ (\emph{lower left panel}). For the two most-massive stellar mass samples, $q_0$ correlates with the dark matter halo mass. For the two samples with the least-massive galaxies, an anti-correlation is observed (\emph{lower right panel}). $q_0$ is normalised by the standard deviation of this parameter for the 500 galaxies in each sample.}
\label{Fig90_PCAScaling}
\end{figure*}

\subsection{Relations between the SFH main mode and galaxy properties}\label{PCAScalings}

We have previously shown that the $q_0$-value of the PCA-decomposed
star formation history of a galaxy captures whether its stellar
population forms early or late. We therefore expect correlations
between $q_0$ and other quantities sensitive to the age of stellar
populations. Figure~\ref{Fig90_PCAScaling} shows how $q_0$ for the
different galaxy mass bins relates to the lookback time when half of
the stellar mass is formed ($T_{50}$), the $g-r$ colour at $z=0$, the black hole mass at $z=0$, and the dark matter mass (friends-of-friends) at $z=0$. $q_0$ correlates strongly with $T_{50}$,
which is in agreement with the star formation histories in
Figure~\ref{Fig88_PCA_Histories}. There is an offset between the
normalisation of this correlation between the different samples. This
is partially because the most-massive galaxies form their stars before
the least massive galaxies (see also Figure~\ref{SFR_histories}).

The correlation between $q_0$ and $g-r$ is also expected, since old
stellar populations are expected to be redder than younger stellar
populations. In all the samples there is an extended distribution of the $g-r$
colour, which splits the galaxies into red galaxies (with high values
of $g-r$) with large $q_0$-values and blue galaxies (low values of
$g-r$) with lower $q_0$-values. The division between the blue and red
galaxy populations is for example seen for the most-massive
galaxies, where the red galaxies have $g-r \gtrsim 0.65$. For the
sample containing the most-massive galaxies there is a larger fraction
of red galaxies than for the samples of less massive galaxies, where
the blue population dominates.

The presence of a red and blue population of galaxies that form their
stars late and early, respectively, is consistent with the \emph{red
  and blue cloud}, which are connected through the \emph{green valley},
established from observations of galaxies
\citep[e.g.][]{2001AJ....122.1861S, 2007ApJ...665..265F,
  2014MNRAS.440.2810B, 2014MNRAS.440..889S}. Galaxies in the blue
cloud are known to be actively star-forming, whereas the red cloud
consists of passive galaxies that are most likely quenched by
feedback processes. In Illustris, there is a correlation between the
star formation history and the black hole mass at $z=0$ for the three
most-massive bins (\emph{lower left panel}, Figure~\ref{Fig90_PCAScaling}),
which suggests that the galaxies that form their stars
early are quenched by AGN feedback, although other interpretations of the correlation cannot be ruled out by the analysis presented here. For the $10^9 \msun$ galaxies,
such a trend is not evident in the simulation, which is expected because the black
hole feedback models in Illustris do not significantly affect galaxies in this mass range.

The panel that shows the dark matter halo mass, $M_\text{halo}$, at $z=0$ versus $q_0$ reveals that these two variables are correlated for the samples with the two highest stellar masses. Furthermore, galaxies that form their stars late (low $q_0$) have halo masses that are close to $10^{12}\msun$, whereas galaxies that form stars early have higher halo masses. This is expected within the framework in which stars are most likely to form in halos of mass $10^{12} \msun$. For the samples with $M_*=10^9\msun$ and $10^{10}\msun$ a weak anti-correlation between $q_0$ and $M_\text{halo}$ is present. The anti-correlation most likely arises because a halo mass of $10^{12}\msun$ is never reached for these galaxies, so they form stars most efficiently at low redshift, when the dark matter halo mass is as close as possible to $10^{12}\msun$.

\begin{figure*}
\centering
\includegraphics[width = 0.8 \textwidth]{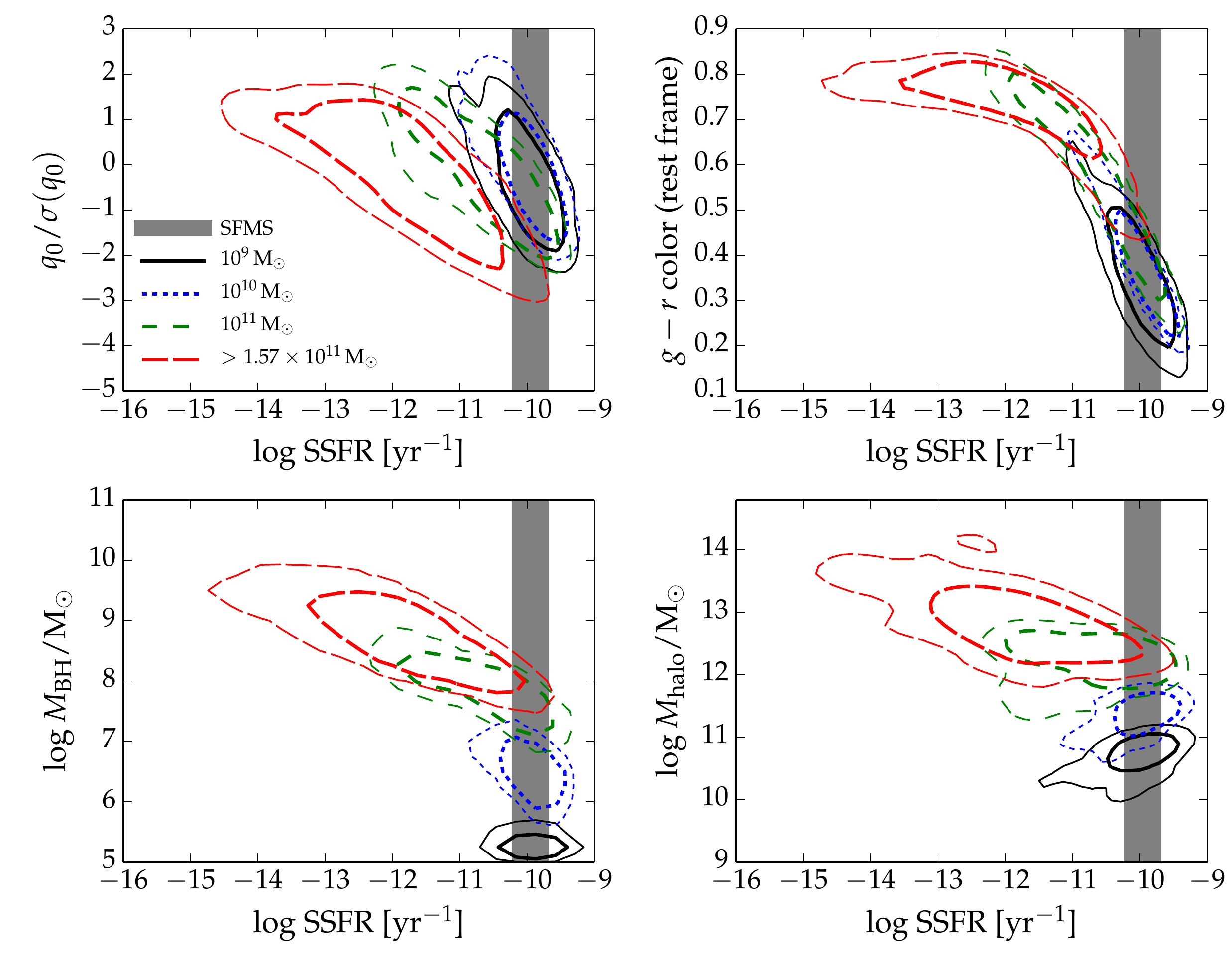}
\caption{This figure shows how various physical quantities behave across the star formation main sequence, which is shown as a grey shaded area (it indicates the median SSFR $\pm 1\sigma$). The confidence intervals and the various physical quantities are the same as in Figure~\ref{Fig90_PCAScaling}. The correlations and anti-correlations in the figure show that the spread in star formation histories (quantified with $q_0$), dark matter halo masses and black hole masses likely are responsible for producing scatter in the star formation main sequence.}
\label{Fig94_SSFR_Scalings}
\end{figure*}

\section{What processes contribute to the scatter in the star formation main sequence?}\label{WhatDrivesScatter}

After having studied the halo growth and star formation histories of individual galaxies, we will now return to the topic of the scatter in the SFMS. In Section~\ref{SFR_indicator}, we showed that the scatter decreased from 0.27 dex to 0.22 dex when calculating the scatter from a 500 Myr-averaged SFR instead of the instantaneous SFR. The scatter in the SFR over timescales of less than 500 Myr is therefore $\simeq$0.16 dex (computed as $\sqrt{0.27^2-0.22^2}$). This allows distinguishing between scatter induced by short-timescale variations ($<500$ Myr) and long-timescale variations ($>500$ Myr), where the latter mode is dominant in Illustris.

In Section~\ref{PCAScalings}, we showed that the principal component, $q_0$, of a galaxy's SFH correlates (or is anti-correlated) with the dark matter halo mass, the black hole mass, and the galaxy colour. In Figure~\ref{Fig94_SSFR_Scalings}, we show how these quantities behave for galaxies above and below the SFMS in different mass bins (we again use the stellar mass bins from Table~\ref{table:MassRanges}). Essentially all galaxies with a SSFR enhanced more than $1\sigma$ above the SFMS have negative $q_0$ values, which implies that they formed their stars later than typical galaxies of the same stellar mass. The galaxies that are $1\sigma$ or more below the SFMS in the two low-mass bins all have positive $q_0$, which implies that these galaxies formed their stars in the early Universe. Taking these considerations into account, it is not surprising that galaxies above the SFMS have bluer $g-r$ colors than galaxies below it.

The decline of black hole mass with SSFR also shows that black hole quenching could be an important mechanism for determining whether a galaxy is in the upper or lower part of the main sequence. Differences in dark matter halo masses among different galaxies also produce	 scatter in the SFMS, since galaxies with $M_\text{halo}\simeq 10^{12}\msun$ (where gas is turned most efficiently into stars) tend to have elevated SSFRs. The halo mass is mainly determined by the long-timescale evolution of a galaxy (i.e. it is relatively insensitive to baryonic processes), so the halo evolution mainly contributes to the long-timescale variation ($>500$ Myr) of a galaxy's SFH. The short-timescale variability is likely due to short-timescale variability in the stellar feedback, gas accretion and black hole activity.


\section{Discussion}\label{Discussion}

\subsection{The normalisation of the star formation main sequence}

In Section \ref{Sec:StarformingPopulation}, we demonstrated that in
Illustris, star-forming galaxies exhibit a tight, approximately linear
correlation between the SFR and stellar mass, as is observed for real
galaxies. Although the normalisation agrees well with observations at $z=0$ and $z=4$, the normalisation in Illustris is less than that observed for intermediate redshifts
$z \sim 1-2$. A
similar problem of under-predicting the normalisation of the main
sequence relation at $z\simeq 2$ has previously been noted by several
authors \citep[e.g.,][]{2007ApJ...670..156D, 2008MNRAS.385..147D,
  2009ApJ...705..617D}, and it appears to be a generic problem for
cosmological hydrodynamical simulations and semi-analytical
models. Thus, it is worthwhile to consider potential solutions to this
problem and discuss whether it is indeed a serious problem.

\subsubsection{Potential theoretical solutions}

In Illustris and other cosmological hydrodynamical simulations, the
SFR-stellar mass relation is a result of the correlation of gas inflow
and outflow rates with halo mass \citep[e.g.][]{Dave2011, Dave2012,
  Dekel2013}. Thus, inaccurate gas net accretion rates would lead to
an incorrect normalisation of the SFR-stellar mass relation.
\citet{2009ApJ...705..617D} compared the growth rate of galaxies in
observations with semi-analytical models \citep{2008MNRAS.384....2G}
and argued that a time-varying IMF cannot resolve the issue of  too
low normalisations at $1\lesssim z\lesssim 2$. Instead,
\citeauthor{2009ApJ...705..617D} suggest that the discrepancy may be
due to the simplified schemes for gas accretion used in
semi-analytical models. \citet{2014arXiv1403.1585M} identified the
same problem of a too low normalisation of the main sequence at $z=2$
in semi-analytical models. They showed that a modification of the
timescale over which gas ejected by feedback is reincorporated into
galaxies can help to fix this problem for galaxies with $M_* \lesssim
10^{10} \msun$ (a modification of the reincorporation timescale was
also studied by \citealt{2013MNRAS.431.3373H}), but it cannot solve
the problem for more-massive galaxies.

Concerns regarding gas inflow rates are less applicable to the
Illustris simulation than to semi-analytical models because in
Illustris gas accretion is explicitly treated using an accurate
hydrodynamics algorithm \citep[e.g.][]{2012MNRAS.424.2999S,2012MNRAS.425.3024V,2012MNRAS.425.2027K,
  2012MNRAS.423.2558B}.  Still, it is possible that the gas cooling
rates in Illustris are systematically offset. For example, if the
phase structure or metal content of the hot halo gas is incorrect, gas
inflow rates could end up being inaccurate \citep{2013MNRAS.429.3353N,
  Hayward2014a}.

Outflows are also important for setting the normalisation of the
SFR-stellar mass relation, but they must still be treated with
sub-resolution models even in state-of-the-art large-volume
cosmological hydrodynamical simulations. One can easily imagine
altering the normalisation of the SFR-stellar mass relation by e.g.
modifying the mass-loading factor of stellar winds. However, the
parameters of the sub-resolution feedback models used in Illustris
were determined by requiring the simulation to match observations such
as the $z = 0$ stellar mass function. Thus far, attempts to tune the
parameters to reconcile the discrepancy with which we are concerned
here without breaking these constraints were unsuccessful
\citep{2014MNRAS.438.1985T}. Nevertheless, this does not preclude the
possibility of solving the discrepancy through the use of
more sophisticated feedback models.

The EAGLE simulation \citep{2014arXiv1407.7040S} is a large-scale cosmological galaxy formation simulation with a feedback model different from the one used in Illustris. A feature of the model in EAGLE is that the wind velocity is untied from the local dark matter velocity dispersion. The EAGLE simulation also has a too-low normalisation of the SFMS at $z=1-2$ \citep{2014arXiv1410.3485F}. They suggest that a burstier star formation model can help solve this problem. The idea is that the SSFR of star-forming galaxies can shift to larger values if galaxies typically form their stars at higher SFRs. However, this requires that galaxies increase the time they spend being passive (i.e. below the SFMS). This suggestion is in good agreement with our finding in Section~\ref{Sec:StarformingPopulation} that the star formation rate has to be further decoupled from dark matter accretion rate in order to produce the correct normalisation of the SFMS at $z=2$. Examples of burstier feedback models are presented in \citet{2013arXiv1311.2073H} and \citet{2010Natur.463..203G}.

\subsection{The paucity of starbursts in Illustris}\label{LowStarburstFraction}

The small number of strong starbursts (i.e.~galaxies that lie
significantly above the star formation main sequence) identified in
Illustris is another discrepancy in addition to the disagreement in
the normalisation of the SFR-stellar mass relation discussed above. As
shown in Fig.~\ref{Fig_StarburstFraction}, at all redshifts
considered, 2.5$\sigma$ outliers from the SFR-stellar mass relation
contribute at most a few per cent of the total SFR density in massive
galaxies ($M_* \ga 10^{10} M_{\odot}$), and this contribution is a
factor of at least a few less than what is inferred from
observations. Furthermore, by combining scalings derived from
idealised merger simulations with a semi-empirical model,
\citet{2010MNRAS.402.1693H} estimated that at all redshifts,
merger-induced starbursts account for $\sim5-10$ per cent of the SFR
density of the Universe, which is also at odds with the Illustris
results.

One possible reason for the relative shortage of starbursts in
Illustris is its kiloparsec-scale spatial resolution. Although the
comparison of the low- and high-resolution runs presented in
Figure~\ref{Fig_StarburstFraction} suggests that the fraction of starbursts is converged, it is possible that this conclusion will change if the resolution is increased significantly, as is often done using zoom-in simulations of galaxies \citep[e.g.;][]{2011ApJ...742...76G,2014MNRAS.437.1750M}.
It is for example possible that the spatial resolution in both the high- and low-resolution Illustris runs is insufficient to resolve the tidal
torques that drive starbursts in mergers. Indeed, the compact sizes
(of order 100 pc) of the starbursting regions in local ULIRGs
\citep[e.g.][]{Sakamoto2008, Engel2011} support this
interpretation. Furthermore, examination of the individual star
formation histories presented in Fig.~\ref{SFR_histories_outliers}
indicates that the galaxies' star formation was burstier at high
redshift ($z \ga 2$).  Because a fixed comoving softening yields finer
resolution at higher redshifts, this observation may also indicate
that resolution is the reason for the lack of starbursts in Illustris.

The relatively stiff equation of state of the
\citet{2003MNRAS.339..289S} sub-resolution ISM model may also serve to
suppress starbursts in Illustris. However, in higher-resolution
idealised merger simulations that employed the same ISM model, SFR
elevations of an order of magnitude or more have been found
\citep[e.g.][]{Cox2008, Hayward2011, Hayward2014a, 2012ApJ...746..108T}. 
Furthermore, the resulting SFR elevations are sufficient to match the observed
interaction induced SFRs of close-pair galaxies in SDSS \citep{Scudder2012,2013MNRAS.433L..59P}.
We note however that galaxy merger simulations with much softer equations
of state show still stronger (albeit shorter) SFR increases
\citep[e.g.][]{Mihos1996}.  Also, the elevation depends sensitively on
the mass ratio and orbital parameters of the merger
\citep[e.g.][]{DiMatteo2007, DiMatteo2008, Cox2008}. Thus, the
sub-resolution ISM model may contribute to the suppression of
merger-driven starbursts in Illustris but is unlikely to be the sole
reason. But mergers are not the only mechanism that can drive
starbursts: it has been suggested that violent disk instability is
also an important channel for driving starbursts
\citep[e.g.][]{Dekel2009,Ceverino2010,Cacciato2012,Porter2014}. Such
events may be suppressed in Illustris because of the sub-resolution
ISM model.

Finally, it is possible that the IR luminosity-inferred star formation
rates of extreme outliers from the main sequence are overestimates of
the true SFR. For highly obscured galaxies, at the peak of the
starburst, simulations suggest that older stellar populations
\citep{2014arXiv1402.0006H} and obscured AGN (L. Rosenthal et al., in
preparation) may contribute to the IR luminosity and thus cause the
SFR to be overestimated by a factor of a few.  These effects are
almost certainly not significant for the bulk of the star-forming
galaxy population and thus should not affect the normalization of the
SFR-stellar mass relation, but, if they are relevant for any real
galaxies, it will likely be those galaxies that are well above the
galaxy main sequence. Consequently, these effects may explain the
tension between observations and Illustris demonstrated in
Section~\ref{LesserRoleStarbursts}.

\section{Conclusions} \label{sec:conclusions}

In this work, we have examined the star formation main sequence and
individual star formation histories of galaxies in the Illustris
simulation.  Our main findings can be summarized as follows:

\begin{itemize}
\item The normalisation of the star formation main sequence is
  consistent with the observations at $z=0$ and $z=4$. At intermediate
  redshifts, $z\sim 2$, the normalisation is significantly lower than
  reported in observations. There is also a slight tension
  between the slope of the star formation main sequence for low-mass
  galaxies, for which Illustris predicts an approximately mass-independent
  specific SFR, whereas observations indicate that the specific SFR is a
  decreasing function of stellar mass. We speculate that 
  more-sophisticated feedback models are required in order to properly recover the
  observed slope and normalisation of the SFMS.

\item The scatter in the star formation main sequence in Illustris is $\simeq 0.27$
dex, which is consistent with observations. Most of the scatter, 0.22 dex, comes from
long-timescale ($\ga 500$ Myr) variability effects; this scatter is likely driven by differences
in dark matter accretion histories of galaxies of a given stellar mass, which can affect the
resulting SFR in multiple ways. A scatter of 0.16 dex originates from variations on timescales
shorter than 500 Myr. This short-timescale variation is likely driven by short-timescale variations
in the gas accretion rate and stellar and black hole feedback.

\item The highest star formation efficiency is found in halos with masses
  of $10^{12}\msun$, and the largest contribution to the star formation
  rate density comes from galaxies in halos with masses of
  $10^{11.5}-10^{12.5} \msun$ at $z\lesssim 4$. At higher redshift,
  the dominant contribution originates in less-massive halos because
  of the declining abundance of halos with high masses. The stellar
  masses of the galaxies contributing most strongly to the global star
  formation rate density lie in the range $10^{10}-10^{11}\, \msun$,
  which is in agreement with observational constraints on this peak
  mass. Another result is that galaxies with stellar masses above $\simeq 10^{11} \msun$ that reside in $\sim10^{12}\msun$ halos form their stars later
  than galaxies in more-massive halos. This is a consequence of star formation being most efficient in $\sim10^{12}\msun$ halos.

\item We have studied the dominant modes and the time variability of
  individual star formation histories with a principal component
  analysis, finding that the characteristic timescale of star
  formation fluctuations in the simulation is $500$ Myr. Another result of this
  principal component analysis is that many features of a galaxys star formation
  history can be described by the leading principal component. 
  This leads us to suggest that star formation histories based on one or several
  principal components can be useful when fitting the spectral energy distribution of observed galaxies.

\item Compared to observations, there is a paucity of strong starbursts
  in Illustris, as evidenced by the small number of galaxies that lie
  significantly above the star formation main sequence. This is likely
  caused in part by a lack of spatial resolution in the cosmological
  simulation, but it may also reflect the relatively stiff equation of
  state model used in Illustris' subgrid model for the regulation of
  star formation in the ISM.
\end{itemize}

In future cosmological simulations of galaxy formation, the relative
frequency of starbursts may well turn out to be an important
constraint that informs about adequate models for the ISM. In
Illustris, starbursts are presumably damped in intensity and stretched
in time by the stiff ISM model, without much affecting the stellar
mass and the structural properties of merger remnants. If a
significant number of starbursts are triggered by disk instabilities,
which are suppressed by the ISM model used in Illustris, the resulting discrepancies could
potentially be more significant. Refined simulation models will be
necessary to clarify this question.

\section*{Acknowledgments}
We thank Sune Toft, Peter Behroozi, Nicholas Lee and David Sanders for useful discussions. The Dark Cosmology Centre is funded by the Danish National
Research Foundation. CCH is grateful to the Klaus Tschira Foundation and the Gordon and Betty Moore Foundation for financial support.
VS acknowledges support by the European Research Council under ERC-StG
EXAGAL-308037.

\def\aj{AJ}
\def\araa{ARA\&A}
\def\apj{ApJ}
\def\apjl{ApJ}
\def\apjs{ApJS}
\def\apss{Ap\&SS}
\def\aap{A\&A}
\def\aapr{A\&A~Rev.}
\def\aaps{A\&AS}
\def\mnras{MNRAS}
\def\nat{Nature}
\def\pasp{PASP}
\def\aplett{Astrophys.~Lett.}
\def\physrep{Physical Reviews}

\footnotesize{
\bibliographystyle{mn2e}
\bibliography{ref}
}

\end{document}